\documentclass[aps,pra,twocolumn,superscriptaddress,longbibliography]{revtex4}
\usepackage{graphicx}
\usepackage{dcolumn}
\usepackage{bm}
\usepackage{physics}
\usepackage{amsmath}
\usepackage{amssymb}
\usepackage{array}
\usepackage{soul}
\usepackage{color}
\usepackage{xcolor} 
\newcolumntype{P}[1]{>{\centering\arraybackslash}p{#1}}
\usepackage[colorlinks=true, breaklinks=true, linkcolor=blue, citecolor=blue, urlcolor=blue]{hyperref}

\newcommand{\unbc}{Department of Physics, University of Northern British Columbia, Prince George, BC V2N 4Z9 Canada}

\newcommand{\pks}{Max Planck Institute for the Physics of Complex Systems, N\"othnitzer Str.~38, 01187 Dresden, Germany}

\begin{document}

\title{Controlling charge and spin currents through nonreciprocal dissipative processes}
\date{\today}

\begin{abstract}
We investigate the generation and control of both charge and spin currents via nonreciprocal dissipative mechanisms
in a two-dimensional spinful fermionic atom quantum system with broken inversion and time-reversal symmetries. 
Within the Gorini-Kossakowski-Sudarshan-Lindblad master equation formalism and using an approach based on the time-dependent generalized Gibbs ensemble,
we identify in the weak dissipative coupling regime the minimal set of nonreciprocal jump operators required to induce charge and
spin currents and to control both their direction and magnitude. We find that in the presence of finite tunneling,
Rashba coupling and magnetic field, the combine application of two jump operators nonreciprocally coupling each spin species
to a different spatial direction of motion is sufficient to generate both types of current. Furthermore, by tuning the degree
of nonreciprocity of the jump operators we modify the dominant transport mechanism from spin to charge. 
Finally, we checked that this nonreciprocal current generation mechanism is robust to the application of dephasing noise
as even in the presence of this additional dissipative process the steady-state occupation distributions for
the quasiparticle modes of the Hamiltonian remains non-trivial, an essential requirement to obtain non-zero currents.
\end{abstract}

\author{Catalin-Mihai Halati}
\affiliation{\pks}
\author{Jean-S\'ebastien Bernier}
\affiliation{\unbc}
\maketitle

\section{Introduction}

How to exert control over the dynamical properties of quantum materials via non-equilibrium schemes is a
question of paramount importance in quantum science \cite{BasovHsieh2017, MivehvarRitsch2021, delaTorreSentef2021, LandiSchaller2022, SchlawinSentef2022,
  KennesRubio2023, DefenuPappalardi2024, SiebererDiehl2025, Chiriaco2025}. In particular, a major challenge
resides in the development of methods to induce non-equilibrium dynamics in many-body systems while maintaining quantum coherence.
To harness genuine quantum effects, addressing this issue is essential. One promising approach to tackle this problem
has been to couple, in a controlled manner, quantum systems to dissipative processes to
stabilize complex quantum properties \cite{DiehlZoller2008, KrausZoller2008, VerstraeteCirac2009, WeimerBuchler2010, MuellerZoller2012, BernierKollath2013}. 
This dissipative engineering approach has been very successful in designing setups to realize
steady states with non-trivial properties. For example, this approach has been used to engineer topological states in fermionic matter
\cite{BardynDiehl2012, BardynDiehl2013, BudichDiehl2015, SheikhanKollath2016, MivehvarPiazza2017} and even to
tailor the dynamical properties of interacting many-body systems \cite{SciollaKollath2015, MacieszczakGarrahan2016, KingMorigi2024, HalatiKollath2025}.
Recently, the realm of possibility offered by dissipative engineering has significantly grown due to its nascent
connection to the field of quantum active matter \cite{SiebererDiehl2025, FruchartVitelli2026}. By taking inspiration from
classical active matter, the design of nonreciprocal dissipative couplings breaking detailed balance has
lead to the emergence of quantum collective effects unknown in equilibrium situations \cite{MetelmannClerk2015, KeckFazio2018, YamamotoKawakami2020, WanjuraNunnenkamp2020, AdachiKawaguchi2022, YamagishiObuse2024, TakasanKawaguchi2024, LeeClerk2024, ZelleDiehl2024, BrighiNunnenkamp2024, KhassehHeyl2025, NadolnyBrunelli2025, BelyanskyClerk2025, AntonovLowen2025, HanaiTazai2025, PennervonOppen2025, Halati2025, ShiBukov2026, MarcheMazza2026, GipoulouxSchiro2026, SoaresSchiro2026}.
Interesting examples include inducing persistent currents without the application of a gauge
field \cite{KeckFazio2018}, realizing entanglement phase transitions without stochasticity \cite{LeeClerk2024}
quantum flocks \cite{KhassehHeyl2025} and dissipative phase transitions \cite{SoaresSchiro2026}, addressing the
interplay of exceptionality and critical fluctuations \cite{ZelleDiehl2024}, and stabilizing currents via dissipative
optical cavity \cite{Halati2025}.

Given the current rapid progress toward a more comprehensive understanding of
quantum active matter, an appealing idea would be to develop ``active matter inspired'' schemes to maintain quantum
coherence in non-equilibrium systems. In closed systems, a striking example of quantum coherence is given by the stabilization
of persistent currents circulating along one-dimensional rings \cite{ButtikerLandauer1983}.
This phenomenon occurs in systems with broken time-reversal symmetry and is reflected in the system quantum state.
The state non-trivial structure is detected by charge current operators as they are probing antisymmetric combinations
of the off-diagonal entries of the corresponding density matrix. In ultracold atomic systems, for example, broken time-reversal
symmetry can be achieved using artificial magnetic or gauge fields \cite{DalibardOehberg2011, GoldmanSpielman2014}. For many-body
systems beyond one-dimensional chains, the search of current carrying systems has led to the discovery of 
complex chiral phases in square \cite{OrignacGiamarchi2001, AtalaBloch2014} and triangular \cite{HalatiGiamarchi2023, MolinelliHouck2026}
ladder geometries, and to the development of Hall response probes for strongly interacting
phases \cite{BuserGiamarchi2021, CitroOrignac2025, HalatiGiamarchi2025, ZhouFallani2023}. Due to recent advances,
measuring local charge currents is now even possible for both ultracold atoms in optical
lattices \cite{ImpertroAidelsburger2024, ImpertroAidelsburger2025}
and superconducting circuits \cite{DuMa2024, MolinelliHouck2026}.

For dissipative quantum systems, most works propose to generate charge currents by coupling reservoirs to the boundaries of a system
(see \cite{LandiSchaller2022} and references therein) with much fewer works introducing uniform dissipative couplings within the bulk
to induce charge currents \cite{KeckFazio2018, YamamotoKawakami2020, ZhengCooper2016, Halati2025}.
In addition to the charge degrees of freedom, spins can also lead to non-trivial transport properties. In fact, the dynamics of spin currents is
the central focus of spintronics \cite{ZuticDasSarma2004, SpinCurrentbook2017, MaekawaSaitoh2023}. In this field, a
key issue relates to conversion processes between charge and spin currents, with the notable example of the spin Hall
effect \cite{Hirsch1999, MurakamiZhang2003, SinovaMacDonald2004, SinovaJungwirth2015}. A typical system to study the interplay
of charge and spin currents is the two-dimensional electron gas in the presence of Rashba
spin-orbit coupling \cite{BychkovRashba1984, ManchonDuine2015, LangeMaekawa2021, FujimotoMaekawa2021}.

\begin{figure}[!hbtp]
\centering
\includegraphics[width=0.45\textwidth]{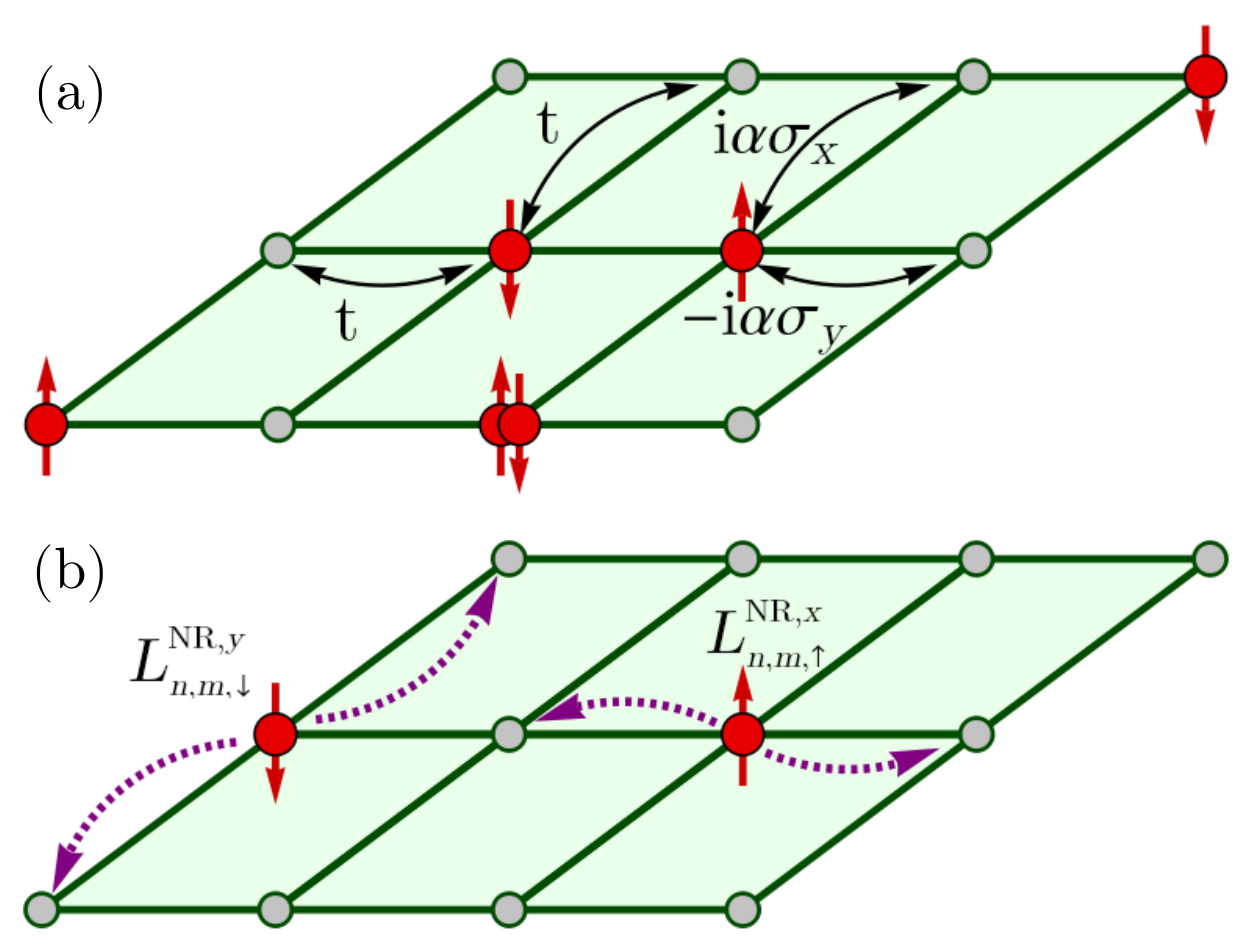}
\caption{Sketch of the model considering spinful fermions in a two-dimensional lattice. The fermionic particles
  experience Hamiltonian, Eq.~\eqref{eq:Hamiltonian_1}, and dissipative processes, Eq.~\eqref{eq:Lindblad}.
  In panel (a), we sketch the coherent tunneling processes with amplitude $t$ which are spatially isotropic and do not change the spin
  state of the particles, and the coherent spin-orbit Rashba coupling of amplitude $\alpha$ for which the tunneling
  is associated with a spin flip process for which the phase depends on the spatial direction (see Sec.~\ref{sec:setup}).
  In panel (b), we sketch the action of two of the dissipative channels considered, Eq.~\eqref{eq:jump_operators}. The jump operators
  depicted act differently on the two spin states, with $L^{\mathrm{NR},x}_{n,m,\uparrow}$ representing a correlated tunneling
  process of the $\uparrow$ state in the $x$-direction, while $L^{\mathrm{NR},y}_{n,m,\downarrow}$ a correlated tunneling process
  of the $\downarrow$ state in the $y$-direction. The combination of the two jump operators leads to a nonreciprocal
  dissipative process reminiscent of spin-orbit coupling because of their directionality and phase, however they do not involve
  spin flips.}
\label{fig:sketch}
\end{figure}

In this article, we investigate the generation of charge and spin currents via
nonreciprocal dissipative mechanisms considering a two-dimensional system of spin-orbit coupled spinful fermionic
particles. We show that the interplay between the coherent Hamiltonian terms and the dissipative
nonreciprocal tunneling processes leads to both finite charge and spin currents. 
Interestingly, the magnitude and direction of both currents can be controlled by tuning
the phase of the dissipative tunneling processes correlating the direction of motion of the particles to their internal
spin states. We further show that the presence of a magnetic field term in the Hamiltonian is crucial
for the emergence of currents. Additionally, we investigate the competition between the nonreciprocal dissipative channels
and reciprocal dephasing noise demonstrating that currents are also present in this regime. We determine the long-time,
steady state, behavior of the dissipative system by employing the time-dependent generalized Gibbs ensemble approach valid
in the weakly dissipative limit. This approach relies on the identification of the Hamiltonian
conserved quantities \cite{LangeRosch2017, LangeRosch2018, LenarcicRosch2018}.

The article is structured as follows: In Sec.~\ref{sec:model}, we describe the model considered in this work,
discussing the Hamiltonian terms, introducing the nonreciprocal dissipative channels and computing the eigenmodes
of the Hamiltonian. In Sec.~\ref{sec:currents}, we derive the currents observables, which are the main focus of our work,
defining the charge and spin currents using U($1$) and SU($2$) gauge transformations and discussing the
dissipative contributions to the currents. The method we employ to obtain the long-time behavior of the
open quantum system, based on time-dependent generalized Gibbs ensembles is presented in Sec.~\ref{sec:method}.
Our results are presented in Sec.~\ref{sec:results}, discussing the charge currents in Sec.~\ref{sec:chargecurrents}
and the spin currents in Sec.~\ref{sec:spincurrents}. We provide a deeper understanding on the necessary
ingredients for generating finite currents in Sec.~\ref{sec:contributions} and analyze the robustness
of the currents to noise in Sec.~\ref{sec:dephasing}. We conclude with Sec.~\ref{sec:conclusion}.

\section{Model \label{sec:model}}

\subsection{Hamiltonian \label{sec:setup}}

We consider spinful fermionic atoms confined to a two-dimensional square lattice, experiencing Rashba
spin-orbit coupling, as sketched in Fig.~\ref{fig:sketch}(a). The Hamiltonian of the system is given by
\begin{align} 
\label{eq:Hamiltonian_1}
H=H_{\text{kin}} +&H_\text{SOC}+H_{\text{mag}} \\
H_{\text{kin}}=-t \sum_{n,m=1}^L& \big(\vec{c}_{n,m}^{\,\dagger} \vec{c}_{n+1,m} +\vec{c}_{n+1,m}^{\,\dagger} \vec{c}_{n,m}\nonumber\\
&+\vec{c}_{n,m}^{\,\dagger} \vec{c}_{n,m+1}+\vec{c}_{n,m+1}^{\,\dagger} \vec{c}_{n,m} \big), \nonumber\\
H_\text{SOC}= i\alpha \sum_{n,m=1}^L &\big(\vec{c}_{n,m}^{\,\dagger} \sigma_x \vec{c}_{n,m+1}-\vec{c}_{n,m+1}^{\,\dagger} \sigma_x \vec{c}_{n,m}\nonumber\\
&-\vec{c}_{n,m}^{\,\dagger} \sigma_y \vec{c}_{n+1,m} +\vec{c}_{n+1,m}^{\,\dagger} \sigma_y \vec{c}_{n,m} \big),\nonumber \\
H_{\text{mag}}=h  \sum_{n,m=1}^L& \vec{c}_{n,m}^{\,\dagger} \sigma_z \vec{c}_{n,m}, \nonumber 
\end{align}
where the spinor vectors $\vec{c}_{n,m}^{\,\dagger} = (c_{n,m,\uparrow}^\dagger, c_{n,m, \downarrow}^\dagger)$ 
and $\vec{c}_{n,m} = (c_{n,m,\uparrow}, c_{n,m, \downarrow})^t$ consist of fermionic annihilation and creation operators
of the particles at position $\vec{r}=(n,m)$ and spin state $\sigma\in\{\uparrow,\downarrow\}$. 
The total atom number is $N=\sum_{n,m} \vec{c}_{n,m}^{\,\dagger} \vec{c}_{n,m}=\sum_{n,m} (n_{n,m,\uparrow}+n_{n,m, \downarrow})$
within a two-dimensional square lattice of size $L\times L$. The particle density is given by $\rho_\mathrm{density} = N/(2L^2)$ such
that a fully filled lattice, $N_{\uparrow} = N_{\downarrow} = L^2$, corresponds to $\rho_\mathrm{density} = 1$.
$H_{\text{kin}}$ describes the tunneling along the two directions of the lattice with amplitude $t$. The
Rashba spin-orbit coupling is modeled by $H_\text{SOC}$ of strength $\alpha$.
The Zeeman magnetic field with magnitude $h$ distinguishes between the spin states and is captured by $H_{\text{mag}}$.
$\sigma_x$, $\sigma_y$ and $\sigma_z$ are the Pauli spin-$1/2$ matrices. We consider periodic boundary conditions in
both spatial directions. For clarity, we also write the terms of the Hamiltonian $H$ in an explicit form
with respect to the spin state indices
\begin{align} 
\label{eq:Hamiltonian_2}
H_{\text{kin}}=-t &\sum_{n,m,\sigma} \big(c_{n,m,\sigma}^{\dagger} c_{n+1,m,\sigma} +c_{n+1,m,\sigma}^{\dagger} c_{n,m,\sigma}\\
&~~+c_{n,m,\sigma}^{\dagger} c_{n,m+1,\sigma}+c_{n,m+1,\sigma}^{\dagger} c_{n,m,\sigma} \big), \nonumber\\
H_\text{SOC}= i\alpha &\sum_{n,m} \big(c_{n,m,\uparrow}^{\dagger} c_{n,m+1,\downarrow}+c_{n,m,\downarrow}^{\dagger} c_{n,m+1,\uparrow}\nonumber\\
&~~-c_{n,m+1,\uparrow}^{\dagger} c_{n,m,\downarrow}-c_{n,m+1,\downarrow}^{\dagger} c_{n,m,\uparrow}\nonumber\\
&~~+ic_{n,m,\uparrow}^{\dagger} c_{n+1,m,\downarrow}-ic_{n,m,\downarrow}^{\dagger} c_{n+1,m,\uparrow} \nonumber\\
&~~-ic_{n+1,m,\uparrow}^{\dagger} c_{n,m,\downarrow}+ic_{n+1,m,\downarrow}^{\dagger} c_{n,m,\uparrow} \big),\nonumber \\
H_{\text{mag}}=h & \sum_{n,m} \big(c_{n,m,\uparrow}^{\dagger}  c_{n,m,\uparrow}-c_{n,m,\downarrow}^{\dagger}  c_{n,m,\downarrow}\big). \nonumber 
\end{align}
In the above expression, we can observe the difference between the tunneling terms in $H_{\text{kin}}$ which do not
act on the spin degrees of freedom, while in $H_\text{SOC}$ the tunneling process is accompanied by a spin flip
with a direction-dependent $\pi/2$ phase.

\subsection{Dissipative processes \label{sec:dissipation}}

In this work we aim to induce charge and spin currents in the model defined in Eq.~(\ref{eq:Hamiltonian_1}) by
employing dissipative couplings. As we consider Markovian dissipative processes, the dynamics of the density
matrix $\rho$ is given by the Gorini-Kossakowski-Sudarshan-Lindblad (GKSL) master
equation \cite{GoriniSudarshan1976, Lindblad1976, CarmichaelBook, BreuerPetruccione2002} 
\begin{align}
\label{eq:Lindblad}
\pdv{t} \rho &=\mathcal{L} \rho=-\frac{i}{\hbar} \left[ H, \rho \right]  +\mathcal{D}(\rho),~\text{with} 	\\
\mathcal{D}(\rho)&= \frac{1}{2}\sum_{j}\Gamma_j\left(2L_j\rho L_j^\dagger-L^\dagger L_j \rho-\rho L_j^\dagger L_j  \right), \nonumber
\end{align}
with Liouvillian $\mathcal{L}$ capturing both the Hamiltonian evolution and the dynamics due to the dissipator $\mathcal{D}(\rho)$
consisting of dissipative processes of strength $\Gamma_j$ for the corresponding jump operator $L_j$. 

Inspired by quantum active matter models in which nonreciprocal dissipative channels are employed,
we consider the following jump operators
\begin{align} 
  \label{eq:jump_operators}
L^{\text{NR},x}_{n,m,\sigma}&=c^\dagger_{n-1,m,\sigma} c_{n,m,\sigma}+e^{i\phi} c^\dagger_{n+1,n,\sigma} c_{n,m,\sigma}, \\
L^{\text{NR},y}_{n,m,\sigma}&=c^\dagger_{n,m-1,\sigma} c_{n,m,\sigma}+e^{i\phi} c^\dagger_{n,m+1,\sigma} c_{n,m,\sigma}, \nonumber
\end{align}
with strength $\Gamma_\text{NR}$, uniform throughout the lattice. These operators realize a correlated hopping
dissipative process, sketch in Fig.~\ref{fig:sketch}(b).
Moreover, if we only employ the jump operators $L^{\text{NR},x}_{n,m,\uparrow}$ and $L^{\text{NR},y}_{n,m,\downarrow}$, we
realize a dissipative analogue of the spin-orbit coupling by nonreciprocally coupling each spin species
with a direction of motion; however, by contrast to spin-orbit coupling, it does not involve spin flips.
We show in Sec.~\ref{sec:results} that the combined action of these two jump operators
generates both charge and spin currents.

Dissipative processes modeled by the jump operators given in Eq.~\eqref{eq:jump_operators} could
in principle be experimentally realized with ultracold atoms in optical lattices by employing Raman transitions
making use of lossy photon modes. This approach would realize couplings of the form
$a_{m,n}^\dagger(\Omega~c^\dagger_{n-1,m,\sigma} c_{n,m,\sigma} + \Omega~e^{i\phi} c^\dagger_{n+1,m,\sigma} c_{n,m,\sigma}) + \text{H.c.}$
with $a_{m,n}^\dagger$ the creation operator for a photon mode labeled by the lattice site
on which it acts. $\Omega$ is the effective strength of the other laser beams used in the Raman
transitions and one should note the presence of a phase difference between the two tunneling processes.
In the limit of very short lifetimes for the photon modes $a_{m,n}$, one can eliminate their
dynamics to obtain effective dissipative channels described by the jump operators detailed
in Eq.~\eqref{eq:jump_operators} \cite{JagerBetzholz2022, BolechGiamarchi2025, SchmitJager2026}.

Furthermore, we check that the currents generated by the action of the nonreciprocal dissipators are robust
under dephasing noise modeled by the following reciprocal jump operator
\begin{align} 
\label{eq:jump_dephasing}
L^d_{n,m}&= n_{n,m,\uparrow}+ n_{n,m,\downarrow}, 
\end{align}
of strength $\Gamma_d$, again uniform throughout the lattice.

\subsection{Diagonalization of the Hamiltonian \label{sec:Hdiag}}

The Hamiltonian described in Eq.~(\ref{eq:Hamiltonian_1}) is non-interacting and quadratic in the fermionic operators.
Thus, we can find the conserved quantities of this system by performing a unitary transformation diagonalizing the Hamiltonian.
As we will investigate the system dynamics using a method based on the time-dependent generalized Gibbs
ensemble (tGGE) \cite{LangeRosch2017, LangeRosch2018, LenarcicRosch2018} (see Sec.~\ref{sec:method}),
identifying the Hamiltonian conserved quantities is essential. As a first step, we perform
a Fourier transformation to momentum space
\begin{align} 
\label{eq:Fourier}
c_{\vec{k},\sigma} = \frac{1}{L} \sum_{n,m=1}^L e^{i \left(k_x n+k_y m\right)} c_{n,m,\sigma},
\end{align}
with the momentum $\vec{k}=(k_x,k_y)$, where $k_{x,y}=\frac{2\pi l}{L},~l=0,\dots, L-1$. 
This transformation renders the kinetic and magnetic field terms of the Hamiltonian in a diagonal form as
\begin{align} 
\label{eq:Hamiltonian_k_1}
H_{\text{kin}}=& \sum_{\vec{k}}\epsilon(\vec{k}) \vec{c}_{\vec{k}}^{\,\dagger} \vec{c}_{\vec{k}},~\text{with}~\epsilon(\vec{k})=-2t\left(\cos k_x+\cos k_y\right) \nonumber\\
H_{\text{mag}}=&~h \sum_{\vec{k}} \vec{c}_{\vec{k}}^{\,\dagger} \sigma_z \vec{c}_{\vec{k}},  
\end{align}
where $\vec{c}^{\,\dagger}_{\vec{k}} = (c_{\vec{k},\uparrow}^\dagger, c_{\vec{k}, \downarrow}^\dagger)$.
However, the spin-orbit coupling Hamiltonian couples the two spin states and in momentum space is given by 
\begin{align} 
\label{eq:Hamiltonian_k_2}
H_{\text{SOC}}=&~2\alpha\sum_{\vec{k}}\vec{c}_{\vec{k}}^{\,\dagger}\left(\sigma_y \sin k_x-\sigma_x \sin k_y\right) \vec{c}_{\vec{k}},  
\end{align}
leading to the following form for the total Hamiltonian
\begin{align} 
\label{eq:Hamiltonian_k_3}
H=& \sum_{\vec{k}}\vec{c}_{\vec{k}}^{\,\dagger}\left[\epsilon(\vec{k}) \mathbb{I}+ 2\alpha\sigma_y \sin k_x-2\alpha\sigma_x \sin k_y+h \sigma_z\right] \vec{c}_{\vec{k}}.  
\end{align}
The Hamiltonian can then be brought in a diagonal form 
\begin{align} 
\label{eq:Hamiltonian_gamma_1}
&H = \sum_{\vec{k}}\left[E_+(\vec{k})\gamma_{\vec{k},+}^{\dagger}\gamma_{\vec{k},+}+E_-(\vec{k})\gamma_{\vec{k},-}^{\dagger}\gamma_{\vec{k},-}\right],\\
&\text{with}~~E_\pm(\vec{k})=\epsilon(\vec{k})\pm\sqrt{4\alpha^2\left(\sin^2 k_x+\sin^2 k_y\right)+h^2}, \nonumber
\end{align}
using the unitary transformation
\begin{align} 
\label{eq:tranformation}
\begin{pmatrix}
c_{\vec{k},\uparrow} \\
c_{\vec{k},\downarrow}
\end{pmatrix}=
\begin{pmatrix}
u_{\uparrow,+}(\vec{k}) & u_{\uparrow,-}(\vec{k})\\
u_{\downarrow,+}(\vec{k}) & u_{\downarrow,-}(\vec{k})
\end{pmatrix}
\begin{pmatrix}
\gamma_{\vec{k},+} \\
\gamma_{\vec{k},-}
\end{pmatrix},
\end{align}
where $\gamma_{\vec{k},+}$ and $\gamma_{\vec{k},-}$ are the fermionic quasiparticle operators representing the eigenmodes of the Hamiltonian. 
The coefficients of the transformation are given by
\begin{align} 
\label{eq:tranformation2}
u_{\uparrow\pm}(\vec{k})&=D_\pm(\vec{k})\left(h\pm\sqrt{\left|b(\vec{k})\right|^2+h^2}\right), \\
u_{\downarrow\pm}(\vec{k})&=D_\pm(\vec{k}) b^*(\vec{k}),\nonumber
\end{align}
with
\begin{align} 
\label{eq:tranformation3}
b(\vec{k})&=-2\alpha\left(\sin k_y+i\sin k_x\right), \\
D_\pm(\vec{k})&=\frac{1}{\sqrt{2}}\left[h\left(h\pm\sqrt{\left|b(\vec{k})\right|^2+h^2}\right)+\left|b(\vec{k})\right|^2\right]^{-1/2}. \nonumber
\end{align}
In addition to this general transformation, it is important to consider separately the transformation for four special momentum values
$\vec{k} = \{(0,0), (0, \pi), (\pi, 0), (\pi,\pi)\}$. For all four cases, the transformation simplifies to
\begin{align} 
\label{eq:tranformation4}
u_{\uparrow +} &= u_{\downarrow -} = 1, \\
u_{\uparrow -} &= u_{\downarrow +} = 0 \nonumber.
\end{align}
This last result also holds in the absence of a magnetic field. Finally, we denote the occupation of each eigenmode, labeled
$\nu=\pm$, as 
\begin{align} 
\label{eq:n_gamma} 
n_\nu(\vec{k})=\left\langle\gamma_{\vec{k},\nu}^{\dagger}\gamma_{\vec{k},\nu}\right\rangle.
\end{align}

\section{Current observables \label{sec:currents}}

In this work, we are interested in generating finite currents in fermionic quantum systems under the action of both
spin-orbit, Eq.~\eqref{eq:Hamiltonian_1}, and dissipative couplings, Eq.~\eqref{eq:jump_operators}.
In such dissipative systems, currents generally consists of two different terms: Hamiltonian and dissipative
currents~\cite{HovhannisyanImparato2019, Halati2025}, where the latter is proportional to the dissipative
coupling strength. In this section, we focus mostly on how the charge and spin currents are defined
for the Hamiltonian part using local U($1$) and SU($2$) gauge transformations~\cite{LeonMillis2008, FujimotoMaekawa2021}.
We first present these transformations before obtaining expressions for both the charge- and spin-current operators
and for the corresponding expectation values highlighting their dependence on two-point correlation functions.
We then briefly explain how to obtain expressions for the dissipative contributions to the currents.

\subsection{\label{sec:transformations} Definitions of the local gauge transformations and of the associated currents}

To derive the charge-current operators on the square lattice, we make use of the
local U($1$) transformation defined as
\begin{align} 
\label{eq:U1_transf_1}
\vec{c}_{n,m} & = U_{n,m}~\vec{\tilde{c}}_{n,m}~~\text{with}~U_{n,m} = e^{i\varphi_{n,m}}
\end{align}
where $\varphi_{n,m}$ is a smooth, slowly varying real function. As $\varphi_{n,m}\ll 1$, the product
of two $U$ transformation matrices can be approximated to
\begin{align} 
\label{eq:U1_transf_2}
U_{n,m}^\dagger U_{n+1,m} &\simeq 1+i (\varphi_{n+1,m}-\varphi_{n,m}) \nonumber \\
                       &\simeq 1+i A^{c,x}_{n,m},  \\
U_{n,m}^\dagger U_{n,m+1} &\simeq 1+i (\varphi_{n,m+1}-\varphi_{n,m}) \nonumber \\
                       &\simeq 1+i A^{c,y}_{n,m},
\end{align}
where we defined the vector potentials corresponding to the U($1$) transformation in the $x$- and $y$-direction as
$A^{c,x}_{n,m}\equiv \varphi_{n+1,m}-\varphi_{n,m}$ and $A^{c,y}_{n,m}\equiv \varphi_{n,m+1}-\varphi_{n,m}$.

By applying this transformation to the Hamiltonian, Eq.~(\ref{eq:Hamiltonian_1}), we can define the charge-current
operators as being linearly coupled to the U($1$) vector potentials. Each operator is obtained
by taking the functional derivative of the transformed Hamiltonian, $\tilde{H}_{\text{U(1)}}$,
with respect to the associated vector potential such that
\begin{align} 
\label{eq:jc_def}
\hat{J}_c^{(n,m); u} &= \left.\frac{\delta \tilde{H}_{\text{U(1)}}}{\delta A^{c,u}_{n,m}}\right\rvert_{\varphi=0},
\end{align}
where the index $u \in\{x,y\}$ labels the charge current direction such that $\hat{J}_{c}^{(n,m);x}$ denotes
the charge-current operator associated with a positive charge current along $x$ from lattice
sites $(n,m)$ to $(n+1,m)$ while $\hat{J}_{c}^{(n,m);y}$ is associated
with a positive charge current along $y$ from sites $(n,m)$ to $(n,m+1)$.

Similarly, we obtain the spin-current operators on the square lattice by exploiting
the SU($2$) gauge transformation defined as
\begin{align} 
\label{eq:SU2_transf_1}
\vec{c}_{n,m} & = V_{n,m} \vec{\tilde{c}}_{n,m}~~\text{with}~V_{n,m} = e^{i\theta\bm{\omega}_{n,m}\cdot\bm{\sigma}}
\end{align}
where $\bm{\sigma}=(\sigma_x,\sigma_y,\sigma_z)$ is the vector of Pauli matrices
and $\bm{\omega}=(\omega_x,\omega_y,\omega_z)$ a vector of smooth real functions, with $|\bm{\omega}_{n,m}|=1$.
As $\theta\ll 1$, the product of two $V$ transformation matrices can be approximated to
\begin{align}
\label{eq:SU2_transf_2}
V_{n,m}^\dagger V_{n+1,m} &\simeq 1 + i \theta\left[(\bm{\omega}_{n+1,m}-\bm{\omega}_{n,m}) \right. \nonumber \\
                       &~~~~~~~~~~~~+\left. \theta(\bm{\omega}_{n,m}\times\bm{\omega}_{n+1,m})\right]\cdot\bm{\sigma} \nonumber \\
                       &\simeq 1+i \mathbf{A}^{s,x}_{n,m} \cdot\bm{\sigma} , \\
V_{n,m}^\dagger V_{n,m+1} &\simeq 1+i \theta \left[(\bm{\omega}_{n,m+1}-\bm{\omega}_{n,m})  \right. \nonumber \\
                       &~~~~~~~~~~~~+\left. \theta(\bm{\omega}_{n,m}\times\bm{\omega}_{n,m+1})\right]\cdot\bm{\sigma} \nonumber \\
                       &\simeq 1+i \mathbf{A}^{s,y}_{n,m} \cdot\bm{\sigma},
\end{align}
where the SU($2$) gauge potentials are defined as 
\begin{align} 
\label{eq:SU2_transf_3}
\mathbf{A}^{s,x}_{n,m} &=\theta(\bm{\omega}_{n+1,m}-\bm{\omega}_{n,m})+\theta^2(\bm{\omega}_{n,m}\times\bm{\omega}_{n+1,m}), \\
\mathbf{A}^{s,y}_{n,m} &=\theta(\bm{\omega}_{n,m+1}-\bm{\omega}_{n,m})+\theta^2(\bm{\omega}_{n,m}\times\bm{\omega}_{n,m+1}).  \nonumber
\end{align}
Using the same approach as before, the spin-current operators are defined as being
linearly coupled to the SU($2$) gauge potentials. The spin-current operators therefore read as
\begin{align} 
\label{eq:js_def}
\hat{J}_{\sigma_\beta}^{(n,m);u} &=\left.\frac{\delta \tilde{H}_{\text{SU(2)}}}{\delta (\mathbf{A}^{s,u}_{n,m})^\beta}\right\rvert_{\theta=0},
\end{align}
where the functional derivative of the transformed Hamiltonian, $\tilde{H}_{\text{SU(2)}}$, is taken with respect to
the $\beta$-component of the gauge potential $\mathbf{A}^{s,u}_{n,m}$ with $\beta\in\{x,y,z\}$. Thus, the local spin-current operators possess
two indexes: the top index $u\in\{x, y\}$ refers to the spatial direction of a positive spin current flow and the bottom index $\sigma_\beta$
refers to the spin direction carried by the current.

\subsection{\label{sec:jc} Charge currents}

Using the gauge transformations and definitions presented in Sec.~\ref{sec:transformations}, we can now
obtain explicit expressions for the charge-current operators and expectation values. Using the U($1$)
transformation, the Hamiltonian, Eq.~(\ref{eq:Hamiltonian_2}), becomes
\begin{align} 
\label{eq:Hamiltonian_U1}
\tilde{H}_{U(1)} =&~ H_{\text{kin}} + H_\text{SOC} + H_\text{mag} \\
& - it \sum_{l,j} A^{c,x}_{l,j} \left(\vec{c}_{l,j}^\dagger \vec{c}_{l+1,j} - \vec{c}_{l+1,j}^\dagger \vec{c}_{l,j} \right) \nonumber \\
& - it \sum_{l,j} A^{c,y}_{l,j}\left(\vec{c}_{l,j}^\dagger \vec{c}_{l,j+1}- \vec{c}_{l,j+1}^\dagger \vec{c}_{l,j} \right) \nonumber \\
& + \alpha \sum_{l,j} A^{c,x}_{l,j}\left(\vec{c}_{l,j}^\dagger \sigma_y \vec{c}_{l+1,j} + \vec{c}_{l+1,j}^\dagger \sigma_y \vec{c}_{l,j} \right) \nonumber \\
& - \alpha \sum_{l,j} A^{c,y}_{l,j}\left(\vec{c}_{l,j}^\dagger \sigma_x \vec{c}_{l,j+1} +  \vec{c}_{l,j+1}^\dagger \sigma_x \vec{c}_{l,j}\right). \nonumber
\end{align}
Following Eq.~(\ref{eq:jc_def}), taking the functional derivative of $\tilde{H}_\text{U(1)}$ with the vector potential $A^{c,u}_{n,m}$ 
leads to the charge-current operator in the $x$-direction \cite{FujimotoMaekawa2021}
\begin{align} 
\label{eq:jc_x}
\hat{J}_{c}^{(n,m);x} = & -it\left(\vec{c}_{n,m}^\dagger \vec{c}_{n+1,m} - \vec{c}_{n+1,m}^\dagger \vec{c}_{n,m} \right) \\
& + \alpha \left(\vec{c}_{n,m}^\dagger \sigma_y \vec{c}_{n+1,m} + \vec{c}_{n+1,m}^\dagger \sigma_y \vec{c}_{n,m} \right), \nonumber
\end{align}
and the charge-current operator in the $y$-direction
\begin{align} 
\label{eq:jc_y}
\hat{J}_{c}^{(n,m);y} = & -it\left(\vec{c}_{n,m}^\dagger \vec{c}_{n,m+1} - \vec{c}_{n,m+1}^\dagger \vec{c}_{n,m} \right) \\
& - \alpha \left(\vec{c}_{n,m}^\dagger \sigma_x \vec{c}_{n,m+1} + \vec{c}_{n,m+1}^\dagger \sigma_x \vec{c}_{n,m} \right). \nonumber
\end{align}
By performing explicitly the products of the spinors with the Pauli matrices, we then obtain the following expectation
values for the charge currents in terms of two-point correlation functions
\begin{align} 
\label{eq:jc_average}
J^x_c &= 2t\left(\Im\langle c_{n,m,\uparrow}^\dagger c_{n+1,m \uparrow}\rangle + \Im\langle c_{n,m,\downarrow}^\dagger c_{n+1,m \downarrow} \rangle\right) \\
      & ~ -2\alpha \left(\Im\langle c_{n,m,\downarrow}^\dagger c_{n+1,m \uparrow}\rangle -\Im\langle c_{n,m,\uparrow}^\dagger c_{n+1,m \downarrow} \rangle\right), \nonumber \\
J^y_c &= 2t\left(\Im\langle c_{n,m,\uparrow}^\dagger c_{n,m+1 \uparrow}\rangle +\Im\langle c_{n,m,\downarrow}^\dagger c_{n,m+1 \downarrow} \rangle\right)  \nonumber \\
& ~ -2\alpha \left(\Re\langle c_{n,m,\downarrow}^\dagger c_{n,m+1 \uparrow}\rangle +\Re\langle c_{n,m,\uparrow}^\dagger c_{n,m+1 \downarrow} \rangle\right). \nonumber
\end{align}
One should note that we have now dropped the site label $(n,m)$ for the charge current expectation values as the system is
translationally invariant. We observe that the charge currents have contributions from both the kinetic terms of the Hamiltonian,
proportional to the tunneling amplitude $t$, and from the Rashba spin-orbit coupling associated to spin flip terms,
proportional to $\alpha$.

\subsection{\label{sec:js} Spin currents}

Similarly, using the SU($2$) gauge transformations and definitions also presented in Sec.~\ref{sec:transformations}, we can
obtain explicit expressions for the spin-current operators and expectation values. In this case, the
Hamiltonian, Eq.~(\ref{eq:Hamiltonian_2}), becomes
\begin{align} 
\label{eq:Hamiltonian_SU2}
& \tilde{H}_\text{SU(2)} \\
&~ = H_{\text{kin}} + H_{\text{mag}} \nonumber \\
&~~- it \sum_{l,j,\beta} (\mathbf{A}^{s,x}_{l,j})^\beta \left(\vec{c}_{l,j}^\dagger \sigma_\beta \vec{c}_{l+1,j} - \vec{c}_{l+1,j}^\dagger \sigma_\beta \vec{c}_{l,j} \right) \nonumber \\
&~~- it \sum_{l,j, \beta} (\mathbf{A}^{s,y}_{l,j})^\beta \left(\vec{c}_{l,j}^\dagger \sigma_\beta \vec{c}_{l,j+1} - \vec{c}_{l,j+1}^\dagger \sigma_\beta \vec{c}_{l,j} \right) \nonumber \\
&~~+ i\alpha \sum_{l,j} \left(\vec{c}_{l,j}^\dagger V^\dagger_{l,j}\sigma_x V_{l,j+1} \vec{c}_{l,j+1}
     - \vec{c}_{l,j+1}^\dagger V^\dagger_{l,j+1}\sigma_x V_{l,j} \vec{c}_{l,j} \right. \nonumber \\
&~~~~~~~~- \left. \vec{c}_{l,j}^\dagger V^\dagger_{l,j}\sigma_y V_{l+1,j} \vec{c}_{l+1,j} 
     + \vec{c}_{l+1,j}^\dagger V^\dagger_{l+1,j}\sigma_y V_{l,j} \vec{c}_{l,j} \right). \nonumber
\end{align}
To simplify Eq.~(\ref{eq:Hamiltonian_SU2}), we make the following change in notation:
$V^\dagger_{n,m}\sigma_\beta V_{n',m'}=V^\dagger_{n,m}\sigma_\beta V_{n,m}V^\dagger_{n,m} V_{n',m'}=\tilde{\sigma}_\beta V^\dagger_{n,m} V_{n',m'}$,
where the transformed Pauli matrices reduce to the usual Pauli matrices in the limit of vanishing $\theta$ such that
$\tilde{\sigma}_\beta\rvert_{\theta=0}={\sigma}_\beta$. Taking into account this change,  Eq.~(\ref{eq:Hamiltonian_SU2}) becomes
\begin{align} 
\label{eq:Hamiltonian_SU2_2}
\tilde{H}_\text{SU(2)} &= H_{\text{kin}} + H_{\text{mag}} \\
&~~- it \sum_{l,j,\beta} (\mathbf{A}^{s,x}_{l,j})^\beta \left(\vec{c}_{l,j}^\dagger \sigma_\beta \vec{c}_{l+1,j} - \vec{c}_{l+1,j}^\dagger \sigma_\beta \vec{c}_{l,j} \right) \nonumber \\
&~~- it \sum_{l,j, \beta} (\mathbf{A}^{s,y}_{l,j})^\beta \left(\vec{c}_{l,j}^\dagger \sigma_\beta \vec{c}_{l,j+1} - \vec{c}_{l,j+1}^\dagger \sigma_\beta \vec{c}_{l,j} \right) \nonumber \\
&~~+ i\alpha \sum_{l,j} \left[\vec{c}_{l,j}^\dagger \tilde{\sigma}_x (1+i\mathbf{A}^{s,y}_{l,j}\cdot\bm{\sigma}) \vec{c}_{l,j+1} \right. \nonumber \\
&~~~~~~~~~~~~~~~~~\left.  - \vec{c}_{l,j+1}^\dagger (1-i\mathbf{A}^{s,y}_{l,j}\cdot\bm{\sigma})\tilde{\sigma}_x \vec{c}_{l,j}\right] \nonumber \\
&~~-i\alpha \sum_{l,j} \left[\vec{c}_{l,j}^\dagger \tilde{\sigma}_y (1+i\mathbf{A}^{s,x}_{l,j}\cdot\bm{\sigma}) \vec{c}_{l+1,j} \right. \nonumber \\
&~~~~~~~~~~~~~~~~~\left.  - \vec{c}_{l+1,j}^\dagger (1-i\mathbf{A}^{s,x}_{l,j}\cdot\bm{\sigma})\tilde{\sigma}_y \vec{c}_{l,j} \right].\nonumber
\end{align}
Then, following Eq.~(\ref{eq:js_def}), taking the functional derivative of $\tilde{H}_{\text{SU(2)}}$ with each component
of the gauge potentials $(\mathbf{A}^{s,u}_{n,m})^\beta$ we obtain expressions for the spin-current operators in the $x$-direction \cite{FujimotoMaekawa2021}
\begin{align} 
\label{eq:js_x}
\hat{J}^{(n,m);x}_{\sigma_\beta} &= -it\left(\vec{c}_{n,m}^\dagger \sigma_\beta \vec{c}_{n+1,m} - \vec{c}_{n+1,m}^\dagger \sigma_\beta \vec{c}_{n,m} \right) \\
&~~~~ +\alpha \left(\vec{c}_{n,m}^\dagger \sigma_y \sigma_\beta \vec{c}_{n+1,m} + \vec{c}_{n+1,m}^\dagger \sigma_\beta\sigma_y \vec{c}_{n,m} \right) \nonumber
\end{align}
and in the $y$-direction
\begin{align} 
\label{eq:js_y}
\hat{J}^{(n,m);y}_{\sigma_\beta} &=-it\left(\vec{c}_{n,m}^\dagger \sigma_\beta \vec{c}_{n,m+1} - \vec{c}_{n,m+1}^\dagger \sigma_\beta \vec{c}_{n,m} \right) \\
&~~~~ -\alpha \left(\vec{c}_{n,m}^\dagger \sigma_x \sigma_\beta \vec{c}_{n+1,m} + \vec{c}_{n,m+1}^\dagger \sigma_\beta\sigma_x \vec{c}_{n,m} \right). \nonumber
\end{align}
Once again we can perform explicitly the products of the spinors with the Pauli matrices and obtain expectation values for the spin currents
in terms of two-point correlation functions. The spin currents in the $x$-spatial direction are
\begin{align} 
\label{eq:js_average_x}
&J^x_{\sigma_x} = 2t\left(\Im\langle c_{n,m,\uparrow}^\dagger c_{n+1,m \downarrow}\rangle +\Im\langle c_{n,m,\downarrow}^\dagger c_{n+1,m \uparrow} \rangle\right) \nonumber \\
&~ +2\alpha \left(\Im\langle c_{n,m,\uparrow}^\dagger c_{n+1,m \uparrow}\rangle -\Im\langle c_{n,m,\downarrow}^\dagger c_{n+1,m \downarrow} \rangle\right), \\
&J^x_{\sigma_y} = -2t\left(\Re\langle c_{n,m,\uparrow}^\dagger c_{n+1,m \downarrow}\rangle -\Re\langle c_{n,m,\downarrow}^\dagger c_{n+1,m \uparrow} \rangle\right) \nonumber \\
&~ +2\alpha \left(\Re\langle c_{n,m,\uparrow}^\dagger c_{n+1,m \uparrow}\rangle +\Re\langle c_{n,m,\downarrow}^\dagger c_{n+1,m \downarrow} \rangle\right),  \nonumber\\
&J^x_{\sigma_z} = 2t\left(\Im\langle c_{n,m,\uparrow}^\dagger c_{n+1,m \uparrow}\rangle -\Im\langle c_{n,m,\downarrow}^\dagger c_{n+1,m \downarrow} \rangle\right) \nonumber \\
&~ -2\alpha \left(\Im\langle c_{n,m,\uparrow}^\dagger c_{n+1,m \downarrow}\rangle +\Im\langle c_{n,m,\downarrow}^\dagger c_{n+1,m \uparrow} \rangle\right), \nonumber
\end{align}
while the spin currents in the $y$-spatial direction are
\begin{align}
\label{eq:js_average_y}
 &J^y_{\sigma_x} = 2t\left(\Im\langle c_{n,m,\uparrow}^\dagger c_{n,m+1 \downarrow}\rangle +\Im\langle c_{n,m,\downarrow}^\dagger c_{n,m+1 \uparrow} \rangle\right) \nonumber \\
&~ -2\alpha \left(\Re\langle c_{n,m,\uparrow}^\dagger c_{n,m+1 \uparrow}\rangle +\Re\langle c_{n,m,\downarrow}^\dagger c_{n,m+1 \downarrow} \rangle\right), \\
&J^y_{\sigma_y} = -2t\left(\Re\langle c_{n,m,\uparrow}^\dagger c_{n,m+1 \downarrow}\rangle -\Re\langle c_{n,m,\downarrow}^\dagger c_{n,m+1 \uparrow} \rangle\right) \nonumber \\
&~ +2\alpha \left(\Im\langle c_{n,m,\uparrow}^\dagger c_{n,m+1 \uparrow}\rangle -\Im\langle c_{n,m,\downarrow}^\dagger c_{n,m+1 \downarrow} \rangle\right),  \nonumber \\
&J^y_{\sigma_z} = 2t\left(\Im\langle c_{n,m,\uparrow}^\dagger c_{n,m+1 \uparrow}\rangle -\Im\langle c_{n,m,\downarrow}^\dagger c_{n,m+1 \downarrow} \rangle\right) \nonumber \\
&~ +2\alpha \left(\Re\langle c_{n,m,\uparrow}^\dagger c_{n,m+1 \downarrow}\rangle -\Re\langle c_{n,m,\downarrow}^\dagger c_{n,m+1 \uparrow} \rangle\right). \nonumber
\end{align}
Here again we dropped the site label $(n,m)$ for the spin current expectation values as the system is translationally invariant.

\subsection{\label{sec:dissj} Dissipative currents}

An alternative approach to derive the current operators is based on the continuity equation, relating the time-evolution
of the local charge or spin densities to currents flowing in and out of a certain site \cite{FujimotoMaekawa2021}.
Using this procedure, the continuity equation can also be used to determine dissipative contributions to the currents.
For the case of the charge currents, the continuity
equation reads 
\begin{align} 
\label{eq:continuity}
\frac{\partial}{\partial t}~\vec{c}_{n,m}^{\,\dagger} \vec{c}_{n,m} &= \left(\hat{J}_c^{(m-1,n);x} + \hat{J}_c^{(m,n-1);y} \right) \\
&~~~ - \left(\hat{J}_c^{(m,n);x} + \hat{J}_c^{(m,n);y} \right) + \hat{Y}_c ,\nonumber
\end{align}
where the two first terms are the Hamiltonian charge-current operators for currents flowing in and out of site $(n,m)$,
as derived in Sec.~\ref{sec:jc}, and $\hat{Y}_c$ is the combination of dissipative charge-current operators.
For example, for the jump operators $L^{\text{NR},x}_{n,m,\sigma}$, Eq.~\eqref{eq:jump_operators}, the
combination of dissipative charge-current operators resulting in charge currents the $x$-direction
would be
\begin{align} 
\label{eq:diss_yc}
\frac{\hat{Y}_c^x}{\hbar~\Gamma_\mathrm{NR}} =
\sum_\sigma & \Big[2~(\hat{n}_{n+1,m,\sigma} + \hat{n}_{n-1,m,\sigma} - 2~\hat{n}_{n,m,\sigma}) \\
&-(e^{i\phi} c_{n+2,m,\sigma}^\dagger c_{n,m,\sigma} + \text{H.c.} )~\hat{n}_{n+1,m,\sigma} \nonumber \\
&+(e^{i\phi}c_{n+1,m,\sigma}^\dagger c_{n-1,m,\sigma} + \text{H.c.} )~\hat{n}_{n,m,\sigma} \nonumber \\
&-(e^{i\phi}c_{n,m,\sigma}^\dagger c_{n-2,m,\sigma} + \text{H.c.} )~\hat{n}_{n-1,m,\sigma} \nonumber\\
&+(e^{i\phi}c_{n+1,m,\sigma}^\dagger c_{n-1,m,\sigma} + \text{H.c.} )~\hat{n}_{n,m,\sigma}\Big]. \nonumber
\end{align}
We observe that $\hat{Y}_c^x$ contains complex four-point correlations related to density-assisted tunneling processes.

For spin currents, one replaces the charge density in Eq.~\eqref{eq:continuity} with the spin density,
$\vec{c}_{n,m}^\dagger \sigma_\beta \vec{c}_{n,m}$, and similarly obtains the Hamiltonian currents derived in Sec.~\ref{sec:js}
and additional dissipative contributions. As an example, the dissipative contribution to the combination of spin current-operators in
the $x$-spatial direction for the spin-$x$ component is
\begin{align} 
\label{eq:diss_yxx}
\frac{\hat{Y}_{\sigma_x}^x}{\hbar~\Gamma_\mathrm{NR}} = &- \sum_\sigma \Big[2~c_{n,m,\sigma}^\dagger c_{n,m,\bar{\sigma}}\\
&+ \hat{n}_{n+1,m,\sigma}~(e^{i\phi}c_{n+2,m,\sigma}^\dagger c_{n,m,\bar{\sigma}}+\text{H.c.})\nonumber\\
&+ \hat{n}_{n-1,m,\sigma}~(e^{i\phi}c_{n-2,m,\sigma}^\dagger c_{n,m,\bar{\sigma}}+\text{H.c.})\nonumber\\
&-e^{-i\phi}(c_{n,m,\sigma}^\dagger c_{n,m,\bar{\sigma}}+\text{H.c.})~c_{n-1,m,\sigma}^\dagger c_{n+1,m,\sigma}\nonumber\\
&-e^{i\phi}(c_{n,m,\sigma}^\dagger c_{n,m,\bar{\sigma}}+\text{H.c.})~c_{n+1,m,\sigma}^\dagger c_{n-1,m,\sigma}\Big],\nonumber
\end{align}
with $\bar{\sigma}$ the opposite spin direction compared to $\sigma$. In this case, we identify four-point correlations
corresponding to density-assisted spin-flip and tunneling processes.

As in this work we are interested in the weakly coupling regime, we focus on the properties of the currents arising
from the Hamiltonian terms, as derived above. In the future, it would be interesting to investigate the dissipative
contributions to the currents when the dissipation strengths are comparable with Hamiltonian energy scales. 

\section{Method \label{sec:method}}

In order to study the role of the dissipative processes described in Sec.~\ref{sec:dissipation} and to compute the non-equilibrium
steady state of the GKSL master equation, Eq.~(\ref{eq:Lindblad}), we use a perturbative approach introduced
in Refs.~\cite{LangeRosch2017, LangeRosch2018, LenarcicRosch2018}. This method, valid in the weak dissipation regime, describes the
system effective dynamics within the space of generalized Gibbs states associated with the symmetries of the Hamiltonian.
Within this regime, the system steady state is approximated by a thermal state whose effective parameters are
determined by solving equations of motion capturing perturbatively the effect of the dissipation. As for the system under study
the Hamiltonian is non-interacting, Eq.~(\ref{eq:Hamiltonian_1}), the thermal state, constructed from the Hamiltonian
conserved quantities, will be given by a generalized Gibbs ensemble (GGE) of the eigenmodes
\cite{LangeRosch2017, LangeRosch2018, LenarcicRosch2018}. This approach was shown to correctly capture the dissipative
dynamics of both spin chains and interacting fermions coupled to non-thermal reservoirs \cite{LangeRosch2018, LenarcicRosch2018} when
benchmarked against small systems. The time-dependent generalized Gibbs ensemble (tGGE) approach has also been used to investigate
current rectification in ultracold atomic systems coupled to spin-dependent dephasing \cite{YamamotoKawakami2020}
and to identify the rapidity distribution as a key quantity to understand the dissipative dynamics of
nonreciprocal XX-spin chains \cite{MarcheMazza2026}. We note that this approach contrasts with other perturbative methods
used for computing long-time states which are based on many-body adiabatic elimination. For the latter, the effective dynamics
is confined to the decoherence free subspace of the dissipator and to parts of the Hamiltonian and is valid for
large dissipative strengths \cite{Garcia-RipollCirac2009, ReiterSorensen2012, Kessler2012, PolettiKollath2013,
  SciollaKollath2015, BernierPoletti2020, HalatiKollath2020b, HalatiKollath2022}.

In the following, we give a short overview of the method based on tGGE following Ref.~\cite{LangeRosch2018} particularized to the
case of a quadratic Hamiltonian, as e.g.~considered in Ref.~\cite{YamamotoKawakami2020}. We close this section with
the derivation of the equations of motion for the jump operators considered in Eq.~(\ref{eq:jump_operators})
and Eq.~(\ref{eq:jump_dephasing}). 

\subsection{\label{sec:eom} Sketch of the tGGE approach}

We split the Liouvillian $\mathcal{L}$, defined in Eq.~(\ref{eq:Lindblad}), into a free part, corresponding to the
Hamiltonian evolution, $\mathcal{L}_0\rho=-i[H,\rho]$ and the perturbation arising from the dissipative
processes, $\mathcal{L}_1\rho=\epsilon\mathcal{D}(\rho)$, with $\epsilon$ small. 
The decoherence free subspace of the free Liouvillian $\mathcal{L}_0$, defined by  $\mathcal{L}_0\rho_0=0$, contains
any mixture of eigenstates of the Hamiltonian $H$. We assume that we can parametrize the decoherence free subspace
of $\mathcal{L}_0$ by generalized Gibbs ensemble states \cite{LangeRosch2017, LangeRosch2018, LenarcicRosch2018}
\begin{align} 
\label{eq:GGE}
\rho_0&=\frac{e^{-\sum_i \lambda_i C_i}}{\Tr\left[e^{-\sum_i \lambda_i C_i}\right]},
\end{align}
with $C_i$ the conservation laws and $\lambda_i$ the corresponding Lagrange parameters. 

In the limit of small $\epsilon$ one can derive the equations of motion for the Lagrange parameters up to
order $\epsilon^2$ \cite{LangeRosch2017, LangeRosch2018, LenarcicRosch2018}
\begin{align} 
\label{eq:eom}
&\frac{\partial}{\partial t}\lambda_i =-\sum_j \left[\chi^{-1}(t)\right]_{ij}\Tr\left[C_j\mathcal{L}_1\rho_0(t)\right],
\end{align}
with $\chi_{ij}=\langle C_i C_j\rangle_0-\langle C_i\rangle_0\langle C_j\rangle_0$ the matrix of two-body connected
correlations evaluated with $\rho_0$. We note that the perturbation parameter $\epsilon$ can be used to rescale time
in Eq.~(\ref{eq:eom}) as $t\to\epsilon t$. Hence, $\epsilon$ does not enter in the calculation of the long-time values
of the parameters of the GGE state $\lambda_i$. However, the GGE steady state obtained using this approach is only
correct in the limit of small $\epsilon$ \cite{LangeRosch2017, LangeRosch2018, LenarcicRosch2018}.

\subsection{\label{sec:H0} Non-interacting Hamiltonian}

In the case of a non-interacting Hamiltonian which can be diagonalized,
$H=\sum_{\vec{q},\nu} E_{\vec{q},\nu} \gamma^\dagger_{\vec{q},\nu} \gamma_{\vec{q},\nu}$
with $\nu=\pm$, Eq.~(\ref{eq:Hamiltonian_gamma_1}), the conserved quantities are the occupations of
each quasiparticle mode $C_{\vec{q},\nu}=\gamma^\dagger_{\vec{q},\nu} \gamma_{\vec{q},\nu}$. Thus, the GGE state is given by
\begin{align} 
\label{eq:GGE_0}
\rho_0&=\frac{1}{Z} e^{-\sum_{\vec{q},\nu} \lambda_{\vec{q},\nu} \gamma^\dagger _{\vec{q},\nu} \gamma_{\vec{q},\nu}}
\end{align}
with the partition function
\begin{align} 
\label{eq:partition}
Z&=\Tr\left[e^{-\sum_{\vec{q},\nu} \lambda_{\vec{q},\nu} \gamma^\dagger _{\vec{q},\nu} \gamma_{\vec{q},\nu}}\right] \nonumber \\
&=\prod_{\vec{q},\nu}\left(1+e^{-\lambda_{\vec{q},\nu} }\right). 
\end{align}
The occupation of the eigenmodes is then expressed as
\begin{align}
\label{eq:occup}
\left\langle \gamma^\dagger_{\vec{q},\nu} \gamma_{\vec{q},\nu} \right\rangle &=\frac{e^{-\lambda_{\vec{q},\nu}}}{1+e^{-\lambda_{\vec{q},\nu}}},
\end{align}
and the connected two-body correlations can also be computed in terms of the parameters of the GGE state as
\begin{align} 
\label{eq:chi}
\chi_{\vec{q},\nu;\vec{k},\mu} &= \left\langle \gamma^\dagger_{\vec{q},\nu} \gamma_{\vec{q},\nu} \gamma^\dagger_{\vec{k},\mu} \gamma_{\vec{k},\mu}\right\rangle-\left\langle \gamma^\dagger_{\vec{q},\nu} \gamma_{\vec{q},\nu}\right\rangle \left\langle \gamma^\dagger _{\vec{k},\mu} \gamma_{\vec{k},\mu}\right\rangle \nonumber \\
&= \delta(\vec{q} - \vec{k})\delta_{\nu,\mu} \frac{e^{-\lambda_{\vec{q},\nu}}}{(1+e^{-\lambda_{\vec{q},\nu}})^2}.
\end{align}
We note that the correlations matrix $\chi_{\vec{q},\nu;\vec{k},\mu}$ is diagonal.

\subsection{\label{sec:diss_eom} Dissipators and Equations of Motion}

To obtain an explicit form for the equations of motion governing the evolution of the
parameters $\lambda_{\vec{q},\nu}$, Eq.~(\ref{eq:eom}), we also need to compute $\Tr[\gamma^\dagger_{q,\nu} \gamma_{q,\nu}\mathcal{L}_1\rho_0]$.
As discussed in Sec.~\ref{sec:dissipation}, we are considering multiple dissipative channels and therefore
write the Liouvillian $\mathcal{L}_1$ as a sum over several dissipators
\begin{align} 
\label{eq:dissipators}
\mathcal{L}_1\rho&=\sum_a\mathcal{D}_a(\rho),\\
\mathcal{D}_a(\rho)&=\frac{\Gamma_a}{2}\sum_{n,m} \left(2L^a_{n,m}\rho L_{n,m}^{a\,\dagger} \right. \nonumber \\
&~~~~~~~~~~~~~~~~~ \left. - L_{n,m}^{a\,\dagger} L^a_{n,m}\rho-\rho  L_{n,m}^{a\,\dagger} L_{n,m}^a\right), \nonumber
\end{align}
where each dissipator contains a single type of jump operator $L^a_{n,m}$ and a sum over the entire lattice.
We can write the jump operators in the basis in which the Hamiltonian is diagonal such that 
\begin{align} 
\label{eq:jump_operator_gspace}
L^a_{n,m}&=\frac{1}{L^2}\sum_{\vec{k},\vec{k}'}\sum_{\mu,\mu'}e^{-i(\vec{k}-\vec{k}')l}K_a^{\mu,\mu'}(\vec{k},\vec{k}')\gamma^\dagger_{\vec{k},\mu} \gamma_{\vec{k}',\mu'},
\end{align}
where the kernel $K_a^{\mu,\mu'}(\vec{k},\vec{k}')$ depends on the transformation diagonalizing the Hamiltonian, Eq.~(\ref{eq:tranformation}), and on
the structure of the jump operator. We keep the form of the kernels general for now and we will later provide explicit expressions for the dissipators
introduced in Sec.~\ref{sec:dissipation}. Combining the expressions found above, we obtain
\begin{align} 
\label{eq:eom2}
&\Tr[\gamma^\dagger_{\vec{q},\nu} \gamma_{\vec{q},\nu}\mathcal{D}_a\rho_0]= \\
&~~~~\frac{\Gamma_a}{2 L^2}\sum_{\vec{k},\vec{k}',\vec{p}}\sum_{\alpha,\alpha',\beta,\beta'}
K_a^{\alpha,\alpha'\,*}(\vec{k},\vec{k}-\vec{p})~K_a^{\beta,\beta'}(\vec{k}'+\vec{p},\vec{k}')\nonumber \\
&~~~~~~~~~~\times\Big\{2\Tr\left[\gamma^\dagger_{\vec{k}-\vec{p},\alpha'} \gamma_{\vec{k},\alpha}\gamma^\dagger_{\vec{q},\nu} \gamma_{\vec{q},\nu}\gamma^\dagger_{\vec{k}'+\vec{p},\beta} \gamma_{\vec{k}',\beta'}\rho_0\right] \nonumber \\
&~~~~~~~~~~~~~~-\Tr\left[\gamma^\dagger_{\vec{q},\nu} \gamma_{\vec{q},\nu}\gamma^\dagger_{\vec{k}-\vec{p},\alpha'} \gamma_{\vec{k},\alpha}\gamma^\dagger_{\vec{k}'+\vec{p},\beta} \gamma_{\vec{k}',\beta'}\rho_0\right] \nonumber \\
&~~~~~~~~~~~~~~-\Tr\left[\gamma^\dagger_{\vec{k}-\vec{p},\alpha'} \gamma_{\vec{k},\alpha}\gamma^\dagger_{\vec{k}'+\vec{p},\beta} \gamma_{\vec{k}',\beta'}\gamma^\dagger_{\vec{q},\nu} \gamma_{\vec{q},\nu}\rho_0\right]\Big\},\nonumber
\end{align}
where we performed the sum over space $\sum_{n,m}$ present in the dissipators and employed the cyclic property of the trace.
By computing the traces over the GGE state $\rho_0$, Eq.~(\ref{eq:GGE_0}), the expression further simplifies to
\begin{align} 
\label{eq:eom3}
&\Tr[\gamma^\dagger_{\vec{q},\nu} \gamma_{\vec{q},\nu}\mathcal{D}_a\rho_0]=\frac{\Gamma_a}{L^2}\sum_{\alpha}\sum_{\vec{p}} \\
&\quad\Bigg[-\left|K_a^{\alpha,\nu}(\vec{q}+\vec{p},\vec{q})\right|^2 \frac{e^{-\lambda_{\vec{q},\nu}}}{(1+e^{-\lambda_{\vec{q},\nu}})(1+e^{-\lambda_{\vec{q}+\vec{p},\alpha}})} \nonumber \\
&\qquad+\left|K_a^{\nu,\alpha}(\vec{q},\vec{q}-\vec{p})\right|^2 \frac{e^{-\lambda_{\vec{q}-\vec{p},\alpha}}}{(1+e^{-\lambda_{\vec{q},\nu}})(1+e^{-\lambda_{\vec{q}-\vec{p},\alpha}})}\Bigg]. \nonumber 
\end{align}
After these manipulations, we obtain the equations of motion for the parameters of the GGE state, Eq.~(\ref{eq:eom}), valid for non-interacting Hamiltonians:
\begin{widetext}
\begin{align} 
\label{eq:eom_lambda}
\frac{\partial}{\partial t} \lambda_{\vec{q},\nu}=-\frac{1}{L^2}\frac{\left(1+e^{-\lambda_{\vec{q},\nu}}\right)^2}{e^{-\lambda_{\vec{q},\nu}}} \sum_{\alpha}\sum_{\vec{p}} \Bigg[&-\left(\sum_a\Gamma_a\left|K_a^{\alpha,\nu}(\vec{q}+\vec{p},\vec{q})\right|^2\right) \frac{e^{-\lambda_{\vec{q},\nu}}}{(1+e^{-\lambda_{\vec{q},\nu}})(1+e^{-\lambda_{\vec{q}+\vec{p},\alpha}})} \\
&+\left(\sum_a\Gamma_a\left|K_a^{\nu,\alpha}(\vec{q},\vec{q}-\vec{p})\right|^2\right) \frac{e^{-\lambda_{\vec{q}-\vec{p},\alpha}}}{(1+e^{-\lambda_{\vec{q},\nu}})(1+e^{-\lambda_{\vec{q}-\vec{p},\alpha}})}\Bigg]. \nonumber 
\end{align}
\end{widetext}

We note that for systems where dissipative channels are particle-conserving, when solving the equations
directly for the steady state $\frac{\partial}{\partial t} \lambda_{\vec{q},\nu}=0$, one also needs to add a condition
for the conservation of the total particle number 
\begin{align} 
\label{eq:particle_number}
N=\sum_{\vec{q},\nu} \langle\gamma^\dagger_{\vec{q},\nu} \gamma_{\vec{q},\nu}\rangle=\sum_{\vec{q},\nu}\frac{e^{-\lambda_{\vec{q},\nu}}}{1+e^{-\lambda_{\vec{q},\nu}}}.
\end{align}
However, when performing the time-evolution of Eq.~(\ref{eq:eom_lambda}), the particle number is automatically conserved if the
dissipative channels considered are particle-conserving. In this work, we used the second approach. Considering a half-filled system,
we use the infinite temperature state for which $\lambda_{\vec{q},\nu} = 0$ for all $\vec{q}$ and $\nu$ as the initial state and
then numerically integrate the set of coupled equations of motion, Eq.~(\ref{eq:eom_lambda}). We obtain all $\lambda_{\vec{q},\nu}$
at every time step and terminate the evolution once they have converged to their steady state values.

\subsection{\label{sec:diss_explicit} Explicit form of the kernels}

For the non-reciprocal jump operators $L^{\text{NR},x}_{n,m,\sigma}$ and $L^{\text{NR},y}_{n,m,\sigma}$, given in Eq.~(\ref{eq:jump_operators}),
of strength $\Gamma_\text{NR}$, the kernels are given by 
\begin{align} 
\label{eq:K3}
K_{\text{NR},x,\sigma}^{\mu,\mu'}(\vec{k},\vec{k}')&=2e^{i\phi/2}\cos\left(k_x-\frac{\phi}{2}\right)u^*_{\sigma,\mu}(\vec{k})u_{\sigma,\mu'}(\vec{k}'), \\
K_{\text{NR},y,\sigma}^{\mu,\mu'}(\vec{k},\vec{k}')&=2e^{i\phi/2}\cos\left(k_y-\frac{\phi}{2}\right)u^*_{\sigma,\mu}(\vec{k})u_{\sigma,\mu'}(\vec{k}'), \nonumber
\end{align}
where $\mu$ and $\mu'$ label the two different kinds of quasiparticle modes: $\gamma_{+}(\vec{k})$ and $\gamma_{-}(\vec{k})$. It is interesting
to note that in these two kernels the non-reciprocal phase is on equal footing with momentum and that the Hamiltonian describing the system is
encoded via the transformation coefficients $u_{\sigma,\mu}(\vec{k})$, Eq.~(\ref{eq:tranformation2}).
For dephasing, with the jump operators $L^d_{n,m}$ and strength $\Gamma_d$, Eq.~(\ref{eq:jump_dephasing}), the kernel is 
\begin{align} 
\label{eq:K1}
K_d^{\mu,\mu'}(\vec{k},\vec{k}')&=u^*_{\uparrow,\mu}(\vec{k})u_{\uparrow,\mu'}(\vec{k}')+u^*_{\downarrow,\mu}(\vec{k})u_{\downarrow,\mu'}(\vec{k}').
\end{align}
Again in this case, the transformation coefficients $u_{\sigma,\mu}(\vec{k})$ carry the information about the Hamiltonian.

\section{Results \label{sec:results}}

The primary objective of this work is to identify the minimal set of jump operators generating non-zero charge and spin currents
while allowing for control over both their direction and magnitude.
Our results presented in this section show that in the presence of finite tunneling, Rashba coupling and magnetic field, the combine
application of the jump operators $L^{\mathrm{NR},x}_{n,m,\uparrow}$ and $L^{\mathrm{NR},y}_{n,m,\downarrow}$ generates finite steady-state
charge and spin currents. Together these two jump operators realize a dissipative equivalent of the spin-orbit couping as
they nonreciprocally couple each spin species to a different spatial direction of motion, but unlike spin-orbit coupling it does
not involve spin flips. As we will show later, the combination of these ingredients leads to non-trivial steady-state occupation distributions
for the quasiparticle modes $\gamma_{\vec{q},+}$ and $\gamma_{\vec{q},-}$ an essential
requirement to obtain non-zero currents.

We present in the following how both the magnitude and direction of the charge and spin currents are controlled by the ratio of
the Rashba coupling to the magnetic field, $\alpha/h$, and the phase, $\phi$.
For the spin degrees of freedom, the dissipative channels are already nonreciprocal in the absence of a phase $\phi$
as each spin state is coupled to a different spatial direction. This should be contrasted to the charge degrees of freedom
as in this case only the presence of a finite phase gives rise to nonreciprocal dissipative couplings.
This important distinction is largely responsible for the different behavior of the spin and charge currents as a
function of $\phi$, discussed in the following sections. After shedding some light on the mechanisms responsible for the generation of currents,
we show that these are robust under the application of
dephasing noise. We consider a half-filled system of size $L \times L = 10 \times 10$ with periodic boundary conditions. 
Initially the system is set to infinite temperature corresponding to $\lambda_{\vec{q},\nu} = 0$ for all $\vec{q}$ and $\nu$
and is further characterized by the absence of both charge and spin currents. 

We note that obtaining a distribution of $\lambda_{\vec{q},\nu}$ different from the one corresponding to the
infinite temperature state would already give rise to a non-equilibrium steady state, however, as we show in the
following only certain non-equilibrium steady states are characterized by finite currents. 
In particular, if we consider the dissipative channels corresponding to the application of the four jump
operators $L^{\mathrm{NR},x}_{n,m,\uparrow}$, $L^{\mathrm{NR},x}_{n,m,\downarrow}$,
$L^{\mathrm{NR},y}_{n,m,\uparrow}$ and $L^{\mathrm{NR},y}_{n,m,\downarrow}$, we find non-trivial values for the
coefficients $\lambda_{\vec{q},\nu}$, but these steady states do not present charge or spin currents.

\subsection{\label{sec:chargecurrents} Charge currents}

\begin{figure}[hbtp!]
  \centering
  \includegraphics[width=0.49\textwidth]{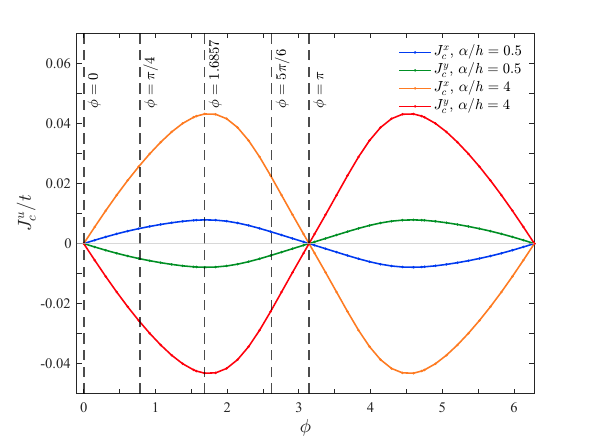}
  \caption{Steady-state charge currents in the $x$-direction, $J^x_c$, and $y$-direction, $J^y_c$, as a
    function of the phase $\phi$ for two ratios $\alpha/h = \{0.5, 4\}$, no dephasing ($\Gamma_\mathrm{d} = 0$) and small $\Gamma_\mathrm{NR}$
    considering a half-filled system of size $L \times L = 10 \times 10$ with periodic boundary conditions.
    The amplitude and direction of the charge currents are strongly dependent on $\phi$. The amplitude of both currents
    also increases with $\alpha/h$. For $0 < \phi < \pi$, $J^x_c$ flows in the positive $x$-direction while $J^y_c$ flows
    in the negative $y$-direction, while the flows are reversed for $\pi < \phi < 2\pi$. Both currents are zero at $\phi = \{0, \pi, 2\pi\}$.}
  \label{fig:chargecurrvsphi}
\end{figure}

\begin{figure}[hbtp!]
  \centering
  \includegraphics[width=0.49\textwidth]{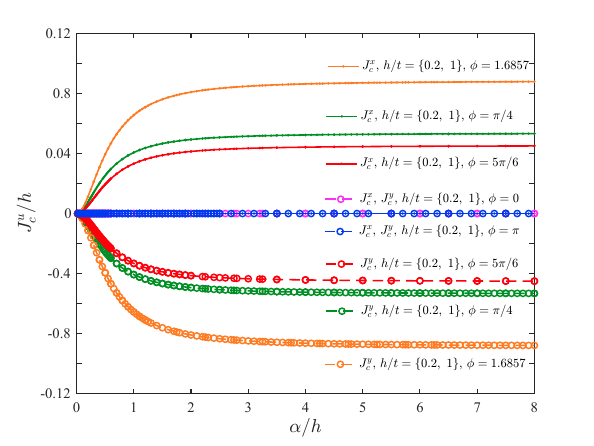}
  \caption{Ratio of the steady-state charge currents in the $x$-direction, $J^x_c/h$, and $y$-direction, $J^y_c/h$, as a function of the
    ratio of the Rashba coupling to the magnetic field, $\alpha/h$, in the absence of dephasing ($\Gamma_\mathrm{d} = 0$),
    small $\Gamma_\mathrm{NR}$, two field strengths $h/t = {0.2, 1}$ and five presentative $\phi$
    values: $0$, $\pi/4$, $1.6875$, $5\pi/6$ and $\pi$. The system is half-filled and of size
    $L \times L = 10 \times 10$ with periodic boundary conditions. Both charge currents are zero when $\phi$ is
    equal to $0$ and $\pi$. For the other $\phi$ considered, when the charge current in $x$ flows
    in the positive direction, the current in $y$ flows in the negative direction.}
  \label{fig:chargecurrvsalpha}
\end{figure}

Starting from the infinite temperature state, where $\lambda_{\vec{q},\nu} = 0$ for all $\vec{q}$ and $\nu$, the system is evolved
using the equations of motion Eq.~\eqref{eq:eom_lambda} corresponding to the combined application of the jump
operators $L^{\mathrm{NR},x}_{n,m,\uparrow}$ and $L^{\mathrm{NR},y}_{n,m,\downarrow}$.
Once the steady state is reached, we evaluate the charge current expectation values as detailed in Eq.~(\ref{eq:jc_average}). 
As illustrated in Fig.~\ref{fig:chargecurrvsphi}, the magnitude and direction of both charge currents $J^x_c$ and $J^y_c$ can be tuned
by changing the nonreciprocal phase $\phi$.  In particular, in the absence of the phase, $\phi=0$, and for $\phi = \pi$,
both charge currents are zero, while the currents become finite for other values of $\phi$. 
Interestingly, by varying the phase $\phi$ one can tune both the magnitude and the direction of the charge currents.
For $0 < \phi < \pi$, the charge current $J^x_c$ flows in the positive $x$-direction whereas $J^y_c$ flows in the negative $y$-direction.
For $\pi < \phi < 2\pi$, the directions of both currents are reversed. 
For all ratios of $\alpha/h$ considered, both charge currents have their maximum value near $\phi \approx 1.6857$.
One should note that in the absence of dissipative processes the ground state of Eq.~(\ref{eq:Hamiltonian_1}) for $h$ and $\alpha$ finite and
at half-filling does not possess charge currents.

For five representative values of the phase, $\phi = \{0, \pi/4, 1.6857, 5\pi/6, \pi\}$, we present in Fig.~\ref{fig:chargecurrvsalpha}
ratios of the charge currents to the magnetic field, $J^x_c/h$ and $J^y_c/h$, as a function of the ratio of the
Rashba coupling to the magnetic field, $\alpha/h$. We obtain that for a given $\phi$, curves corresponding to different magnetic
fields collapse onto a single line highlighting that both $|J^x_c/h|$ and $|J^y_c/h|$ monotonically increase with $\alpha/h$.
This collapse of the curves underscores the processes leading to non-trivial contributions to the currents
which we further discuss in Sec.~\ref{sec:contributions}.

\subsection{\label{sec:spincurrents} Spin currents}

\begin{figure}[hbtp!]
  \centering
  \includegraphics[width=0.49\textwidth]{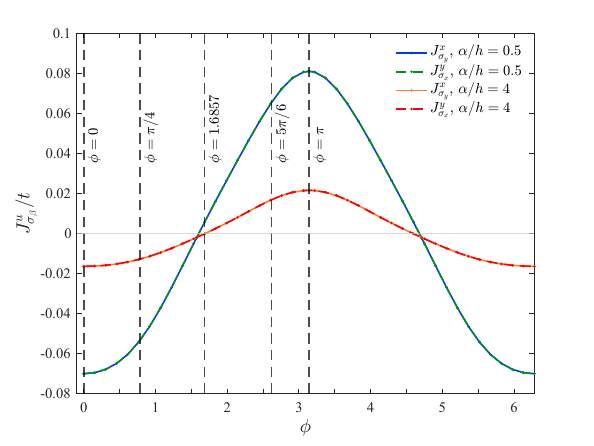}
  \caption{Steady-state spin-$y$ component of the spin current in the $x$-direction, $J^x_{\sigma_y}$, and
    spin-$x$ component of the spin current in the $y$-direction, $J^y_{\sigma_x}$, as a
    function of the phase $\phi$ for two ratios $\alpha/h = \{0.5, 4\}$, small $\Gamma_\mathrm{NR}$
    and $\Gamma_\mathrm{d} = 0$ considering a half-filled system of size
    $L \times L = 10 \times 10$ with periodic boundary conditions. Both spin currents $J^x_{\sigma_y}$
    and $J^y_{\sigma_x}$ have the same value and the direction of both currents can be reversed by changing
    $\phi$.}
  \label{fig:spincurrvsphi}
\end{figure}

\begin{figure}[hbtp!]
  \centering
  \includegraphics[width=0.49\textwidth]{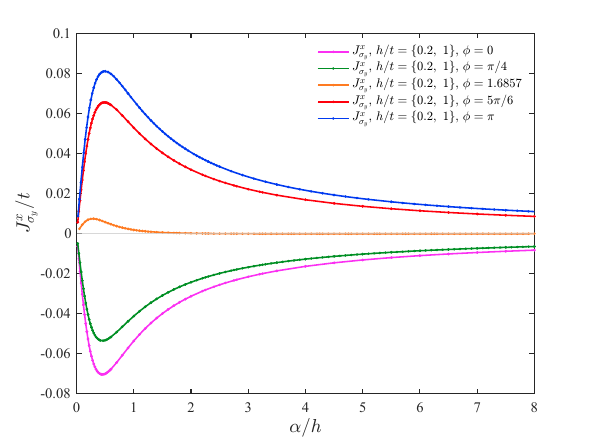}
  \caption{Steady-state spin-$y$ component of the spin current in the $x$-direction as a function of the ratio
    of the Rashba coupling to the magnetic field, $\alpha/h$, in the absence of dephasing ($\Gamma_\mathrm{d} = 0$)
    and small $\Gamma_\mathrm{NR}$, two field strengths $h/t = {0.2, 1}$ and five presentative $\phi$
    values: $0$, $\pi/4$, $1.6875$, $5\pi/6$ and $\pi$. The system is half-filled and of size $L \times L = 10 \times 10$
    with periodic boundary conditions. For $\phi = 1.6875$,
    the spin current goes to zero as $\alpha/h$ increases. For all five values of $\phi$, the spin current presents
    a maximum for $\alpha/h < 1$.}
  \label{fig:spincurrvsalpha}
\end{figure}

As explained in Sec.~\ref{sec:js}, the spin current along the spatial directions $x$ and $y$ each have three components. 
Under the application of the jump operators $L^{\mathrm{NR},x}_{n,m,\uparrow}$ and $L^{\mathrm{NR},y}_{n,m,\downarrow}$, we found
that only two of these six components, $J^x_{\sigma_y}$ and $J^y_{\sigma_x}$, are non-zero
in the steady state. Both these components are equal for the regimes explored in this work.
This contrasts with the
ground state of Eq.~(\ref{eq:Hamiltonian_1}) for $h$ and $\alpha$ finite and at half-filling where $J^x_{\sigma_y} = - J^y_{\sigma_x}$.
In Fig.~\ref{fig:spincurrvsphi}, we show the behavior of the non-zero steady-state spin currents as a function of the phase $\phi$.
While the spin currents also vary with $\phi$, their overall behavior is
different compared to the charge currents (see Fig.~\ref{fig:chargecurrvsphi}).  
As shown in Fig.~\ref{fig:spincurrvsphi}, we observe that already for $\phi = 0$ the spin currents are finite,
in contrast to the charge currents, highlighting that the dissipative coupling between the spin states and the directions of motion
realized by the jump operators $L^{\mathrm{NR},x}_{n,m,\uparrow}$ and $L^{\mathrm{NR},y}_{n,m,\downarrow}$ is already sufficient to
maintain finite spin currents. The spin currents have even their maximal values at $\phi = \{0, \pi, 2\pi\}$, phase values for which the
charge currents are zero. Moreover, the two spin currents are almost zero near $\phi = 1.6857$ where the charge currents are optimal.
Therefore by varying $\phi$, besides the possibility to control the magnitude and sign of both the charge and spin currents,
one can choose values such that solely charge or spin currents are generated.

In Fig.~\ref{fig:spincurrvsalpha}, we show that for a given $\phi$ the spin current $J^x_{\sigma_y}/t$ (which is equal to $J^y_{\sigma_x}/t$)
collapses for different magnetic field values onto a single curve as a function of $\alpha/h$. We further discuss the origin
of this collapse in Sec.~\ref{sec:contributions}. In contrast to the charge currents, for the five representative values
of $\phi$ considered, the spin current rises rapidly for small $\alpha/h$, reach an maximum value at a $\alpha/h$ which depends on $\phi$ and
then slowly drops as the ratio of the Rashba coupling to the magnetic field is further increased. 
For $\phi = 1.6857$, the spin current even quickly reaches vanishing values for $\alpha/h > 1$. Furthermore, for a given $\alpha/h$,
the maximum positive spin current found for $\phi = \pi$ is not equal in magnitude to the minimal negative current found for $\phi = 0$.

\subsection{\label{sec:contributions} Understanding the contributions to the currents}

In this section, we analyze the contribution to the currents stemming from the occupation of each eigenmode $\gamma_{\vec{q},\nu}$.
This will help in understanding the conditions for which finite charge and spin currents emerge and the parameter collapse
shown in Fig.~\ref{sec:chargecurrents} and Fig.~\ref{sec:spincurrents}.
We rewrite the expression of the currents given in terms of the two-point correlation functions, Eq.~(\ref{eq:jc_average}) and
Eqs.~(\ref{eq:js_average_x})-(\ref{eq:js_average_y}), as a function of the occupation of the eigenmodes of the Hamiltonian. 
We consider the examples of the charge current in the $x$-direction, $J^{x}_c$, and the $\sigma_y$ component of the spin current
in the $x$-direction, $J^x_{\sigma_y}$, given by 
\begin{align} 
\label{eq:jc_x_total}
J^{x}_c &= \frac{2t}{L^2}\sum_{\vec{q},\nu}\sin(q_x)n_\nu(\vec{q}) \\
&~~~~-\frac{4\alpha}{L^2}\sum_{\vec{q},\nu}\cos(q_x)\Im\left[u_{\downarrow\nu}^*(\vec{q})u_{\uparrow\nu}(\vec{q})\right]n_\nu(\vec{q}),\nonumber
\end{align}
and
\begin{align} 
\label{eq:js_x_y_total}
J^x_{\sigma_y} &= \frac{4t}{L^2}\sum_{\vec{q},\nu}\sin(q_x)\Im\left[u_{\downarrow\nu}^*(\vec{q})u_{\uparrow\nu}(\vec{q})\right]n_\nu(\vec{q}) \\
&~~~~+\frac{2\alpha}{L^2}\sum_{\vec{q},\nu}\cos(q_x)n_\nu(\vec{q}),\nonumber
\end{align}
where we have used the relation $\left|u_{\uparrow\nu}\left(\vec{q}\right)\right|^2+\left|u_{\downarrow\nu}\left(\vec{q}\right)\right|^2=1$,
which can be derived based on the expressions from Eqs.~\eqref{eq:tranformation2}-\eqref{eq:tranformation3}.

In the following, we first discuss why the currents vanish in the absence of the magnetic field term due to the presence
of a symmetry in the model considered, pinpointing the crucial role of the magnetic field in the generation of currents.
Afterwards, we highlight the relevant contributions to the different type of currents when a magnetic field is present.

\begin{figure}[hbtp!]
  \centering
  \includegraphics[width=0.49\textwidth]{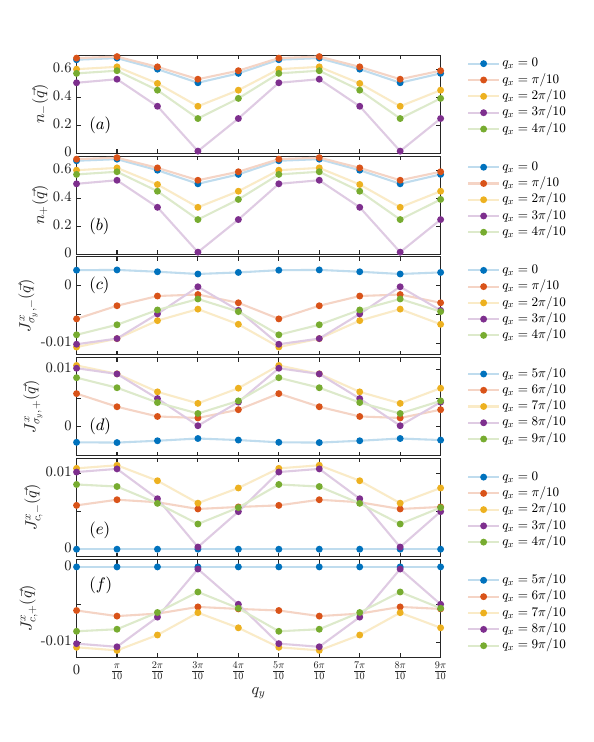}
  \caption{Momentum-dependent contributions to the spin and charge currents due to each mode ($\nu = \{ +, -\}$) for
    $h/t = 0$, $\alpha/t = 0.2$ and $\phi = \pi/4$, considering a half-filled system of size $L \times L = 10 \times 10$ with
    periodic boundary conditions and in the absence of dephasing, $\Gamma_\mathrm{d} = 0$. (a) Occupation
    of each $\gamma_{\vec{q},-}$ mode as a function of momentum, $\vec{q}$. (b) Occupation of each $\gamma_{\vec{q},+}$ mode as function
    of momentum. While these distributions are non-trivial, they are the same for both modes.
    (c) Momentum-dependent contribution to the spin current
    $J^x_{\sigma_y}$ for each $\gamma_{\vec{q},-}$ mode. (d) Momentum-dependent contribution to $J^x_{\sigma_y}$ for
    each $\gamma_{\vec{q},+}$ mode. As $J^x_{\sigma_y,-}(\vec{q}) = -J^x_{\sigma_y,+}(\vec{q}+\pi\vec{e}_x)$, the contributions
    cancel two by two leading to $J^x_{\sigma_y} = 0$. (e) Momentum-dependent contribution to the charge current $J^x_{c}$
    for each $\gamma_{\vec{q},-}$ mode. (f) Momentum-dependent contribution to $J^x_{c}$ for each $\gamma_{\vec{q},+}$ mode. Again as
    $J^x_{c,-}(\vec{q}) = -J^x_{c,+}(\vec{q}+\pi\vec{e}_x)$, one finds $J^x_{c} = 0$.}
  \label{fig:lambdah0}
\end{figure}

\begin{figure}[hbtp!]
  \centering
  \includegraphics[width=0.49\textwidth]{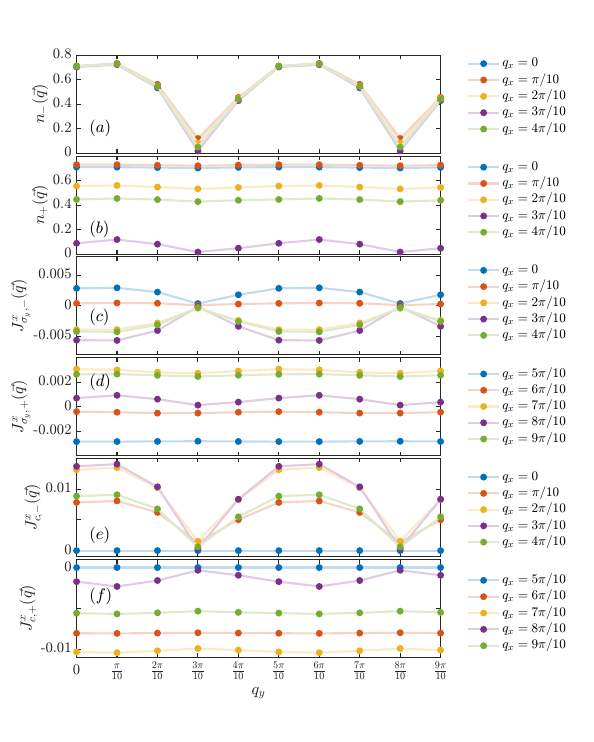}
  \caption{Momentum-dependent contributions to the spin and charge currents due to each mode ($\nu = \{ +, -\}$) for
    $h/t = 1$, $\alpha/t = 0.2$ and $\phi = \pi/4$, considering a half-filled system of size $L \times L = 10 \times 10$ with
    periodic boundary conditions and in the absence of dephasing, $\Gamma_\mathrm{d} = 0$. (a) Occupation
    of each $\gamma_{\vec{q},-}$ mode as a function of momentum, $\vec{q}$. (b) Occupation of each $\gamma_{\vec{q},+}$ mode as function
    of momentum. In contrast to $h/t = 0$, for $h/t = 1$, the mode occupancy distributions are drastically different
    for $\nu = -$ and $\nu = +$. (c) Momentum-dependent contribution to the spin current
    $J^x_{\sigma_y}$ for each $\gamma_{\vec{q},-}$ mode. (d) Momentum-dependent contribution to $J^x_{\sigma_y}$ for
    each $\gamma_{\vec{q},+}$ mode. As the contributions from each modes are cleary different, their addition leads to a non-zero
    spin current $J^x_{\sigma_y}$. (e) Momentum-dependent contribution to the charge current $J^x_{c}$
    for each $\gamma_{\vec{q},-}$ mode. (f) Momentum-dependent contribution to $J^x_{c}$ for each $\gamma_{\vec{q},+}$ mode. Again
    the contributions from each mode are different resulting in a non-zero charge current $J^x_{c}$.}
  \label{fig:lambdah1}
\end{figure}

\subsubsection{In the absence of the magnetic field\label{sec:cont_h0}}

In the absence of a magnetic field, the transformation diagonalizing the Hamiltonian, Eqs.~\eqref{eq:tranformation2}-\eqref{eq:tranformation3},
reduces to $u_{\uparrow+}(\vec{k})=\frac{1}{\sqrt{2}}$, $u_{\uparrow-}(\vec{k})=-\frac{1}{\sqrt{2}}$,
$u_{\downarrow+}(\vec{k})=u_{\downarrow-}(\vec{k})=\frac{b^*(\vec{k})}{\sqrt{2}|b|(\vec{k})}$, with $b(\vec{k})=-2\alpha(\sin k_y+i\sin k_x)$
except for $\vec{k} \in \mathcal{S}$ with $\mathcal{S} = \{(0,0), (0,\pi), (\pi,0), (\pi,\pi)\}$ where Eq.~\eqref{eq:tranformation4} still holds.
This leads to the following expressions for $J^{x}_c$ and $J^x_{\sigma_y}$
\begin{align} 
\label{eq:jc_x_total_h0}
J^{x}_c &=\frac{2t}{L^2}\sum_{\vec{q},\nu}\sin(q_x)n_\nu(\vec{q}) \\
&~~~~+\frac{2\alpha}{L^2}\sum_{\vec{q} \notin \mathcal{S},\nu}\text{sign}(\nu)\frac{\cos(q_x)\sin(q_x)}{\sqrt{\sin(q_x)^2+\sin(q_y)^2}}n_\nu(\vec{q}),\nonumber
\end{align}
\begin{align} 
\label{eq:js_x_y_total_h0}
J^x_{\sigma_y} &=-\frac{2t}{L^2}\sum_{\vec{q} \notin \mathcal{S},\nu}\text{sign}(\nu)\frac{\sin(q_x)^2}{\sqrt{\sin(q_x)^2+\sin(q_y)^2}}n_\nu(\vec{q}) \nonumber\\
&~~~~+\frac{2\alpha}{L^2}\sum_{\vec{q},\nu}\cos(q_x)n_\nu(\vec{q}).
\end{align}

In Fig.~\ref{fig:lambdah0}(a)-(b) we plot the steady-state occupations of the eigenmodes $n_\pm(\vec{q})$ as a
function of the momentum component $q_y$ for several values of the momentum component $q_x$. Two main observations
can be made: (i) $n_-(\vec{q})=n_+(\vec{q})$ and (ii) $n_\pm(\vec{q})=n_\pm(\vec{q}+\pi\vec{e}_{x})=n_\pm(\vec{q}+\pi\vec{e}_{y})$,
for all momentum values. These arise because of the symmetries of the equations of motion, Eq.~\eqref{eq:eom_lambda}, and are
consequences of both the form of the Hamiltonian in the absence of a magnetic field and of the structure of the jump operators.
These symmetry constraints on the steady-state occupations of the eigenmodes have profound
implications for the values of the currents. On one hand, the $\vec{q}\to \vec{q}+\pi\vec{e}_{x}$ symmetry implies that for $J^{x}_c$
the term proportional to $t$, Eq.~\eqref{eq:jc_x_total_h0}, and for $J^x_{\sigma_y}$ the term proportional to $\alpha$, Eq.~\eqref{eq:js_x_y_total_h0}, have
contributions of equal magnitude but different sign for $\vec{q}$ and $\vec{q}+\pi\vec{e}_{x}$.
On the other hand, for $J^{x}_c$ the term proportional to $\alpha$, Eq.~\eqref{eq:jc_x_total_h0}, and for $J^x_{\sigma_y}$ the term proportional to $t$,
Eq.~\eqref{eq:js_x_y_total_h0}, have
different sign for the contributions coming from the $+$ and $-$ eigenmodes.
Thus, these two symmetries determine the absence of charge and spin currents.
This can be observed in Fig.~\ref{fig:lambdah0}(e)-(f) for $J^{x}_c$ and in Fig.~\ref{fig:lambdah0}(c)-(d) for $J^x_{\sigma_y}$, as the individual
momentum contributions to the currents are mostly finite, but they come with opposite signs and same magnitude for the $+$ and $-$ eigenmodes,
leading to vanishing total currents. Similarly the other currents, $J^{y}_c$, $J^y_{\sigma_x}$, discussed in Sec.~\ref{sec:chargecurrents}
and Sec.~\ref{sec:spincurrents}, are also vanishing. Thus, in order to obtain finite currents one needs to break one of the two
symmetries \cite{YamamotoKawakami2020}. In the next section, we show that by adding a magnetic field term to the Hamiltonian, Eq.~\eqref{eq:Hamiltonian_1},
we break the symmetry between the $+$ and $-$ eigenmodes and obtain finite charge and spin currents.

\subsubsection{Finite magnetic fields\label{sec:cont_hnon0}}

For finite magnetic fields, the $\vec{q}\to \vec{q}+\pi\vec{e}_{x/y}$ symmetry of the equations of
motions, Eq.~\eqref{eq:eom_lambda}, is still present. However, the symmetry between the $+$ and $-$  eigenmodes is broken. This can
be seen in the steady-state occupations of the eigenmodes displayed in Fig.~\ref{fig:lambdah1}(a)-(b) for $h/t=1$,
as $n_-(\vec{q})$ and $n_+(\vec{q})$ are clearly different.
Thus the relevant contributions $J^{x}_c$ and $J^x_{\sigma_y}$ are
\begin{align} 
\label{eq:j_x_total_2}
\frac{J^{x}_c}{h}=&-\frac{4}{L^2}\frac{\alpha}{h}\sum_{\vec{q},\nu}\cos(q_x)\Im\left[u_{\downarrow\nu}^*(\vec{q})u_{\uparrow\nu}(\vec{q})\right]n_\nu(\vec{q}), \\
\frac{J^x_{\sigma_y}}{t}=&\frac{4}{L^2}\sum_{\vec{q},\nu}\sin(q_x)\Im\left[u_{\downarrow\nu}^*(\vec{q})u_{\uparrow\nu}(\vec{q})\right]n_\nu(\vec{q}). \nonumber
\end{align}
The individual momentum contribution for $J^{x}_c$ and $J^x_{\sigma_y}$ are shown in Fig.~\ref{fig:lambdah1}(c)-(f),
which in the presence of the magnetic field sum to a finite value. 

In Fig.~\ref{fig:chargecurrvsalpha} from Sec.~\ref{sec:chargecurrents} and Fig.~\ref{fig:spincurrvsalpha} from Sec.~\ref{sec:spincurrents},
we saw that  $J^{x}_c/h$ and $J^x_{\sigma_y}/t$ depend only on the ratio $\alpha/h$ and not separately on
the magnetic field and Rashba coupling strength. Based on Eq.~\eqref{eq:j_x_total_2}, this would imply that the coefficients of
the transformations diagonalizing the Hamiltonian, $u_{\sigma\nu}(\vec{q})$, and the occupations, $n_\nu(\vec{q})$, need to be
function of $\alpha/h$. As the parameters of the Hamiltonian enter the equations of motion, Eq.~\eqref{eq:eom_lambda},
only via the kernels $K^{\alpha,\nu}_a$ which are functions of $u_{\sigma\nu}(\vec{q})$, Eqs.~\eqref{eq:K3}-\eqref{eq:K1},
we need to consider the dependence of $u_{\sigma\nu}(\vec{q})$ on $h$ and $\alpha$.
Based on Eqs.~\eqref{eq:tranformation2}-\eqref{eq:tranformation3}, we can write the coefficients $u_{\sigma\nu}(\vec{q})$ as
\begin{align} 
\label{eq:u_field}
&u_{\uparrow\pm}\left(\vec{q}\right)=D'_\pm\left(\vec{q};\alpha/h\right)\left(1\pm\sqrt{1+\frac{\alpha^2}{h^2}\left|b'\left(\vec{q}\right)\right|^2}\right), \\
&u_{\downarrow\pm}\left(\vec{q}\right)=D'_\pm\left(\vec{q};\alpha/h\right)\frac{\alpha}{h} b'^*\left(\vec{q}\right), \nonumber \\
&\text{with}~D'_\pm\left(\vec{q};\alpha/h\right)=hD_\pm\left(\vec{q}\right), ~b'\left(\vec{q}\right)=\frac{1}{\alpha} b\left(\vec{q}\right). \nonumber
\end{align}
Thus, $u_{\sigma\nu}(\vec{q})$ are functions only of the ratio $\alpha/h$, explaining the behavior discussed in Sec.~\ref{sec:chargecurrents}
and Sec.~\ref{sec:spincurrents}.

\subsection{\label{sec:dephasing} Effect of dephasing}

\begin{figure}[hbtp!]
  \centering
  \includegraphics[width=0.49\textwidth]{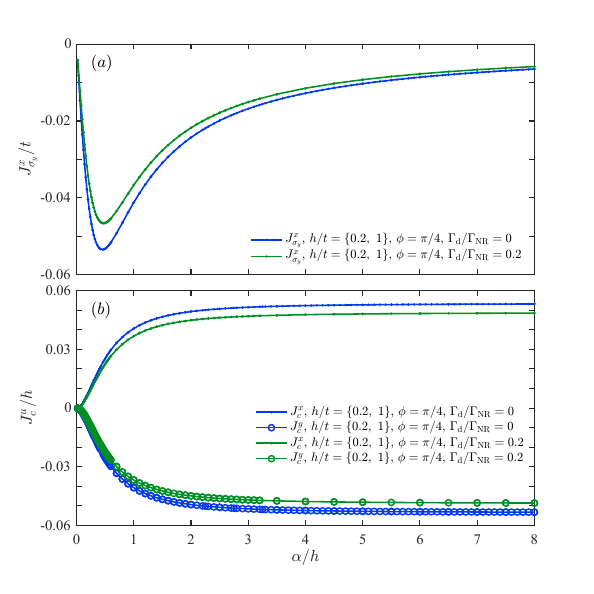}
  \caption{Steady-state spin-$y$ component of the spin current in the $x$-direction $(a)$ and charge currents in both the $x$-
    and $y$-directions $(b)$ as a function of the ratio of the Rashba coupling to the magnetic field for two field strengths
    $h/t = \{0.2, 1\}$, $\phi = \pi/4$. The system is half-filled and of size $L \times L = 10 \times 10$ with periodic boundary
    conditions. Both charge and spin currents are robust to
    dephasing, $\Gamma_\mathrm{d}/\Gamma_\mathrm{NR} = 0.2$.}
  \label{fig:dephasing}
\end{figure}

A non-trivial problem in the field of open quantum system relates to the competition between dissipative channels
that, if taken independently, would drive quantum systems to different steady states. In this section, we investigate
if the steady states generated by the jump operators
$L^{\mathrm{NR},x}_{n,m,\uparrow}$ and $L^{\mathrm{NR},y}_{n,m,\downarrow}$, Eq.~\eqref{eq:jump_operators}, and
characterized by finite charge and spin currents are robust under the action of dephasing noise.
As the dephasing jump operators $L^d_{n,m}$, Eq.~\eqref{eq:jump_dephasing}, are Hermitian, considered alone they
would lead the system to an infinite temperature steady state regardless of the details of the Hamiltonian.

In Fig.~\ref{fig:dephasing}, we show the dependence of the spin and charge currents on $\alpha/h$ in the absence
and presence of dephasing. We obtain for a sizable dephasing strength $\Gamma_\mathrm{d}/\Gamma_\mathrm{NR} = 0.2$ that the
currents are only slightly decreased and that all other features remain present. This shows that the generation and control
of charge and spin currents by employing the nonreciprocal dissipative channels $L^{\mathrm{NR},x}_{n,m,\uparrow}$
and $L^{\mathrm{NR},y}_{n,m,\downarrow}$ are robust even in the presence of dephasing noise.

\section{Conclusions\label{sec:conclusion}}

In summary, we find that, even in the weakly dissipative regime, nonreciprocal processes can be employed
to generate quantum coherent dynamics exemplified by the stabilization of currents. We show that
nonreciprocally coupling each spin species with a direction of motion, making use of the
jump operators $L^{\mathrm{NR},x}_{n,m,\uparrow}$ and $L^{\mathrm{NR},y}_{n,m,\downarrow}$, drives spinful
fermionic particles into non-trivial steady states. These dissipative processes, together with coherent tunneling,
spin-orbit coupling and the application of a magnetic field, lead to the emergence of finite charge
and spin currents. We uncover the minimal ingredients to generate finite currents, highlighting the
crucial role of the magnetic field. Interestingly, the magnitude, direction and even the type of
currents, charge or spin, can be controlled by tuning the phases $\phi$ characterizing the
nonreciprocal action of the dissipators. Furthermore, we show that the generated currents are robust
under the action of dephasing noise. Our work offers exciting novel avenues to build up quantum coherence
in correlated fermionic systems using  nonreciprocal dissipative processes. We foresee that
the nonreciprocal setup proposed could be integrated into cold atom platforms and that the
resulting control over quantum coherence could be assessed using techniques recently developed
to measure current observables from local correlations in cold atomic
systems \cite{ImpertroAidelsburger2024, ImpertroAidelsburger2025}.

\section*{DATA AVAILABILITY}

The supporting data for this article are openly available at Zenodo \cite{datazenodo}.

\section*{ACKNOWLEDGMENTS}

We thank T.~Giamarchi and R.~D.~Soares for fruitful discussions.
We acknowledge support by the Natural Sciences and Engineering Research Council of Canada (NSERC)
[funding references No. RGPIN-2021-04338 and No. DGECR-2021-00359] (J.-S.~B.).

\end{document}